\documentclass[twocolumn]{aastex631}
\usepackage{amsmath}
\usepackage{multirow}

\begin{document}

\title{Atacama Cosmology Telescope: Constraints on the Millimetre Flux of the Crab Pulsar}

\author[0000-0002-1697-3080]{Yilun~Guan}
\affiliation{Dunlap Institute for Astronomy and Astrophysics, University of Toronto, 50 St. George St., Toronto, ON M5S 3H4, Canada}
\footnote{yilun.guan@utoronto.ca}

\author[0000-0002-4478-7111]{Sigurd~Naess}
\affiliation{Institute of Theoretical Astrophysics, University of Oslo, Norway}

\author[0009-0004-5304-3650]{Ian Niebres}
\affiliation{David A. Dunlap Dept of Astronomy and Astrophysics, University of Toronto, 50 St George Street, Toronto ON, M5S 3H4, Canada}
\affiliation{Surge AI, 50 9th Avenue, 5th Floor, New York, NY 10011}

\author[0009-0009-2369-327X]{Louis Branch}
\affiliation{David A. Dunlap Dept of Astronomy and Astrophysics, University of Toronto, 50 St George Street, Toronto ON, M5S 3H4, Canada}

\author[0000-0003-1690-6678]{Adam~D.~Hincks}
\affiliation{David A. Dunlap Dept of Astronomy and Astrophysics, University of Toronto, 50 St George Street, Toronto ON, M5S 3H4, Canada}
\affiliation{Specola Vaticana (Vatican Observatory), V-00120 Vatican City State}

\author[0000-0003-3851-7518]{Hongbo Cai}
\affiliation{Department of Astronomy, School of Physics and Astronomy, Shanghai Jiao Tong University, Shanghai, 200240, China}
\affiliation{Key Laboratory for Particle Astrophysics and Cosmology (MOE)/Shanghai Key Laboratory for Particle Physics and Cosmology, Shanghai, China}

\author[0000-0002-7145-1824]{Allen Foster}
\affiliation{Joseph Henry Laboratories of Physics, Jadwin Hall, Princeton University, Princeton, NJ 08544, USA}

\author[0000-0002-4765-3426]{Carlos~Herv\'ias-Caimapo}
\affiliation{Instituto de Astrof\'isica and Centro de Astro-Ingenier\'ia, Facultad de F\'isica,
Pontificia Universidad Cat\'olica de Chile, Av. Vicu\~na Mackenna 4860, 7820436 Macul, Santiago, Chile}

\author[0000-0002-8816-6800]{John~P.~Hughes}
\affiliation{Department of Physics and Astronomy, Rutgers, the State University of New Jersey, 136 Frelinghuysen Road, Piscataway, NJ 08854-8019, USA}

\author[0000-0002-8149-1352]{Crist\'obal~Sif\'on} 
\affiliation{Instituto de F{\'{i}}sica, Pontificia Universidad Cat{\'{o}}lica de Valpara{\'{i}}so, Casilla 4059, Valpara{\'{i}}so, Chile}

\author[0000-0002-7567-4451]{Edward J. Wollack}
\affiliation{NASA Goddard Space Flight Center, 8800 Greenbelt Road, Greenbelt, MD 20771, USA}

\begin{abstract}
  Millimeter-wave observations of pulsars, while crucial for understanding their emission mechanisms, remain scarce. We demonstrate that high-precision cosmic microwave background (CMB) experiments like the Atacama Cosmology Telescope (ACT), though designed for cosmology, offer a unique capability for such time-domain science due to their high cadence and broad sky coverage in millimeter bands. While previous ACT searches have focused on transients lasting minutes or longer, we develop and validate analysis methods to search for periodic, millisecond-scale transients, a capability not typically associated with CMB experiments. We describe a phase-resolved mapmaking approach, which leverages the known periodicity of the signal to enhance sensitivity and offers advantages in diagnosing systematic errors. We also introduce a template-based fit to the raw data timestreams that independently validate our results. Applying these methods to estimate the millimeter flux of the Crab Pulsar (PSR B0531+21), we derive 95\% confidence upper limits of 4.6\,mJy, 4.4\,mJy, and 20.7\,mJy on the pulsar's period-averaged flux density at 96\,GHz, 148\,GHz, and 225\,GHz, respectively. These constraints fill a gap in our knowledge of the Crab Pulsar's spectral energy distribution, suggesting that it does not significantly flatten or invert at millimeter wavelengths. This work demonstrates the potential for future searches of short-timescale astrophysical phenomena with the next-generation CMB experiments like the Simons Observatory.
\end{abstract}

\section{Introduction} \label{sec:intro}
Millimeter-wave (mm) observations of pulsars provide a critical window into their emission physics, yet this regime remains relatively unexplored. The mechanism responsible for the coherent radio emission of pulsars is still not well understood (see \citealt{Melrose:2017} for a review), and certain models suggest the possibility of a flattening or even an upturn in the pulsar spectrum at mm wavelengths \citep{Kramer:1996,Kramer:1997:7mm}, possibly accompanied by a transition from coherent to incoherent emission. Measurements or upper limits of the mm flux of pulsars can therefore help discriminate between theoretical models. Furthermore, the mm-band offers a cleaner view towards the Galactic center, as it is less affected by the pulse-broadening scatter that hinders radio observations \citep{Cordes:1997,Macquart:2015,Torne:2021}. To our knowledge, 10 pulsars and magnetars have been detected in the mm to date,\footnote{They are: PSR~0329+54, 0355+54, 1929+10, 2020+28, 2021+51 \citep{Wielebinski:1993,Kramer:1997}; PSR B0355+54 \citep{Morris:1997,Torne:2017b}; the Vela Pulsar \citep[PSR~J0834--4511][]{Mignani:2017}; SGR~J1745$-$2900 \citep{Torne:2015,Torne:2017,Torne:2021}; XTE~J1810$-$197 \citep{Camilo:2007,Torne:2020,Torne:2022}; PSR~J1622$-$4950 and 1E~1547.0$-$5408 \citep{Chu:2021,Camilo:2008}.} but it is only recently that these objects have been regularly observed at the shorter end of the mm regime ($\lesssim 7$\,mm). The sole pulsar in this category is the Vela Pulsar, detected by the Atacama Large Millimeter/sub-millimeter Array (ALMA) at 97.5--343.5\,GHz \citep[0.8--3\,mm;][]{Mignani:2017}; additionally, all four magnetars seen in the mm, SGR~J1745$-$2900, XTE~J1810$-$197, PSR~J1622$-$4950 and 1E~1547.0$-$5408, have been detected at various frequencies between 87 and 291\,GHz \citep[1.0--3.4\,mm;][]{Camilo:2007,Torne:2015, Torne:2017,Torne:2020,Torne:2021,Chu:2021,Torne:2022}.

The Crab Pulsar (PSR B0531+21) serves as an excellent benchmark for further pulsar and magnetar measurements in the mm. Not only is its mm flux unknown, but as a young and energetic pulsar with a period of approximately 33~ms and exceptionally well-characterized timing properties \citep{Lyne:1993:crab}, it is an ideal test case for instruments with millisecond temporal resolution. Its emission profile is known to change dramatically at radio frequencies $\gtrsim 5$~GHz \citep{Hankins:2015}, where the main-pulse/inter-pulse structure evolves into a more complex profile. Since the profile at millimeter wavelengths is unknown, a search based on a low-frequency radio template would be inappropriate, motivating a generic, phase-binned search that is robust to uncertainties in the pulse shape. Furthermore, its location within the bright Crab Nebula presents a significant analysis challenge, making it an excellent object for stress-testing our analysis pipelines.

Parallel to these astrophysical motivations, the field of precision measurement of the cosmic microwave background (CMB) has achieved remarkable success over recent decades. Instruments such as the Atacama Cosmology Telescope (ACT) and the South Pole Telescope (SPT) have been instrumental in this progress, producing maps of the CMB at arcminute resolution \citep[e.g.,][]{Chown:2018,ACT:Naess:2025}. With over a decade of millimeter-sky mapping at millisecond-scale temporal resolution, CMB instruments like ACT and SPT provide unique opportunities to explore time-domain phenomena at millimeter (mm) wavelengths, a relatively unexplored area.

The ACT and SPT collaborations have now produced several studies in the area of time domain astronomy. The first search for transients with a ground-based CMB experiment was undertaken by \citet{SPT:Whitehorn:2016} using the SPTpol camera, who analyzed 6,000 hours of observations covering 100 square degrees and identified a possible gamma-ray burst afterglow. \citet{ACT:Naess:2021} serendipitously discovered three mm transients in 3-day maps of ACT maps, coincident in position with nearby stars and likely due to stellar flares; soon thereafter, \citet{SPT:Guns:2021} reported 15 transient detections from a systematic search in SPT data, the majority consistent with stellar flares, but some of possible extragalactic origin. Subsequent systematic transient searches by both collaborations have identified more events: for ACT, \cite{ACT:Li:2023} identified 14 transient events, all likely stellar flares, and more recently, \cite{ACT:Biermann:2025} discovered 34 additional transient events, including one coincident with a classical nova and another associated with a flaring active galactic nucleus; for SPT, \citet{SPT:Tandoi:2024} identified 66 flaring stars with 111 total transient events. \cite{ACT:Hervias-Caimapo:2024} used ACT data to place upper limits on the mm fluxes of extragalactic transients such as gamma-ray burst, tidal-disruption events, and supernovae.

Beyond these astrophysical transients, both ACT and SPT have conducted systematic searches for moving objects in the Solar System, exploiting the fact that the thermal emission at mm wavelengths, which goes as distance from the Earth squared, is stronger than reflected sunlight at shorter wavelengths, going roughly as distance to the fourth power: \cite{ACT:Naess:2021:P9} used ACT data to place new constraints on the existence of a hypothetical `Planet 9' in the outskirts of the Solar System. \citet{SPT:Chichura:2022} and \citet{ACT:Orlowski-Scherer:2024} provided mm flux measurements of three and 170 asteroids in SPT and ACT data, respectively. \citet{SPT:Foster:2025} reported the first detection of thermal emission from artificial satellites at millimeter wavelengths using SPT-3G data.

In addition to ground-based CMB experiments, the \textit{Planck} satellite has also contributed to mm-wave time-domain astronomy through its long-term monitoring of the mm sky, enabled by a scan strategy that allows detection of variability on timescales from days to years. This capability has allowed for systematic studies of AGN variability and polarization, resulting in a catalog of 153 variable compact sources \citep{Rocha:2023}. \citet{Yuan:2016} also reported the detection of millimeter emission from a tidal-disruption event (IGR J12580+0134) in \textit{Planck} data.

While these discoveries have so far focused on events that vary from minutes to days, ACT data are recorded at up to 400\,Hz, presenting an opportunity to probe time-domain phenomena on timescales as short as milliseconds, such as pulses from pulsars, magnetars, rotating radio transients (RRATs), and fast radio bursts (FRBs). ACT, with its frequency coverage at approximately 96 GHz, 148 GHz, and 225 GHz, along with its temporal coverage spanning several years, is therefore a promising instrument for investigating the mm emission of pulsars. Crucially, ACT's near-simultaneous observations across its three bands can constrain the pulsar's spectral index in this largely unexplored region, should a detection be made.

In this work, we explore this potential by presenting two methods for measuring pulsar fluxes using the ACT dataset, though the methods are generalizable to observatories with similar observing strategies like SPT, the Fred Young Submillimeter Telescope (FYST; \cite{CCAT-PrimeCollaboration:2023}), and the Simons Observatory (SO; \citealt{SO:SO_Collaboration:2019,SO:SO_Collaboration:2025}). We focus on the Crab Pulsar as a benchmark to demonstrate these methods.

This paper is organized as follows: In Section~\ref{sec:act}, we provide a description of the ACT instrument and the dataset used in our analysis. Section~\ref{sec:methods} outlines the methods we developed for estimating pulsar flux. In Section~\ref{sec:results}, we present and discuss our findings for the Crab Pulsar. We conclude in Section~\ref{sec:conclusion}.

\section{ACT}
\label{sec:act}

\begin{deluxetable*}{c|cc|cccc}
    \tablecaption{ACT detector array properties.}
        \tablehead{
            Array & 
            Filter 3\,dB\tablenotemark{a} & 
            Sampling Rate\tablenotemark{b} & 
            Band Name & 
            Band Center\tablenotemark{c} & 
            \multicolumn{2}{c}{Number of Detectors\tablenotemark{d}}\\
            & (Hz) & (Hz) & & (GHz) & Total & Live
        }
    \startdata
        \multirow{2}{*}{PA4} & \multirow{2}{*}{121.9} & \multirow{2}{*}{300.5} &
              f150 & 148.3 & 1006 & 475 \\
        & & & f220 & 225.0 & 1006 & 502 \\
        \hline
        \multirow{2}{*}{PA5} & \multirow{2}{*}{141.8} & \multirow{2}{*}{395.3} &
              f090 &  96.5 & 852 & 738 \\
        & & & f150 & 149.2 & 852 & 745 \\ 
        \hline
        \multirow{2}{*}{PA6} & \multirow{2}{*}{141.8} & \multirow{2}{*}{395.3} &
              f090 &  95.3 & 852 & 637 \\
        & & & f150 & 147.8 & 852 & 651 \\
        \hline
    \enddata
    \tablenotetext{a}{A four-pole low-pass filter is applied to the data, acquired at several kHz, before down sampling to the final sampling rate.}
    \tablenotetext{b}{Sampling rate of the recorded TODs.}
    \tablenotetext{c}{Band centers are calculated for a synchrotron spectrum, i.e., $\nu^{-0.7}$. The band center values are taken from \citet{ACT:Hervias-Caimapo:2024}; see that paper for more details.}
    \tablenotetext{d}{The live detector count represents the number of useable detectors out of the total. Of these, on average ${\sim}$70--90\% pass the data cuts, depending on observing conditions. See \citet{ACT:Naess:2021}, the source of these data, for details.}
    \label{tab:array_properties}
\end{deluxetable*}

ACT was a 6-meter off-axis Gregorian telescope situated at an elevation of 5,190 meters on Cerro Toco in Chile's Atacama Desert. The telescope housed three generations of receivers: MBAC (2007--11; \citealt{ACT:Swetz:2011}), ACTPol (2013--2015; \citealt{ACT:Thornton:2016}) and Advanced ACTPol (AdvACT, 2016--22; \citealt{ACT:Henderson:2016}), but only in the era of AdvACT did the survey area include the Crab nebula. AdvACT was equipped with three polarized arrays (PA) of transition-edge sensor (TES) detectors. Each PA was dichroic, meaning that it observed two frequency bands simultaneously; three principal bands were used, centered (for a synchrotron spectrum, $\nu^{-0.7}$) at approximately 96, 148, and 225\,GHz and designated as f090, f150, and f220, and with approximate full width at half maximum beam sizes at these frequencies of 2.0, 1.4 and 1.0\,arcminutes respectively. Table~\ref{tab:array_properties} lists the synchrotron band centers and number of detectors for each array. PA7, with frequency bands centered at 27 and 38\,GHz (f030 and f040) was installed in 2020, replacing PA6, but its data are still being characterized and are not included in this work.

ACT observed by performing scans in azimuth as the sky rotated through its fixed elevation. The TES detectors were read out by a Multi-Channel Electronics (MCE) system at several kilohertz, low-pass filtered and downsampled to about 300\,Hz (PA4) or 395\,Hz (PA5 and PA6) before being recorded to disk. The filter cutoff and exact sampling rates are listed in Table~\ref{tab:array_properties} and further details are given in Appendix~\ref{app:downsampling}. The fundamental data product was thus a collection of 5420 time streams (3716 after 2020, if PA7 is excluded), though the average yield of working detectors was about 55\%, depending on observing conditions \citep{ACT:Naess:2021}. Time streams were stored in approximately 11 minute-long segments; we refer to the bundle of all the detectors' time streams in one of these segments as a TOD.\footnote{TOD stands for `time-ordered data', which is unfortunately ambiguous since this could be taken to refer to a single time stream. It is, however, the established term used in the ACT literature.}

The physical response of the ACT detectors to an impulse can be modeled as $e^{-t/\tau}$, where $t$ is the time and $\tau$ is the detector's time constant. For the AdvACT detectors, the average time constant was $\tau \sim 1\,\mathrm{ms}$ \citep{ACT:Naess:2021}, i.e., faster than the low pass filter and the sampling period. This rapid detector response, coupled with the telescope's absolute timing accuracy of better than 1\,ms, make ms-scale time-domain studies possible with ACT.

In this study, we analyzed observations of the Crab pulsar (PSR B0531+21) from 2017 to 2022. Our dataset consisted of 213, 326, and 113 TODs at f090, f150, and f220, respectively. These data were obtained through a combination of incidental scans during CMB observations and dedicated calibration scans of the Crab Nebula.

\section{Flux Measurements}
\label{sec:methods}

\begin{figure}[t]
  \centering
  \includegraphics[width=\linewidth]{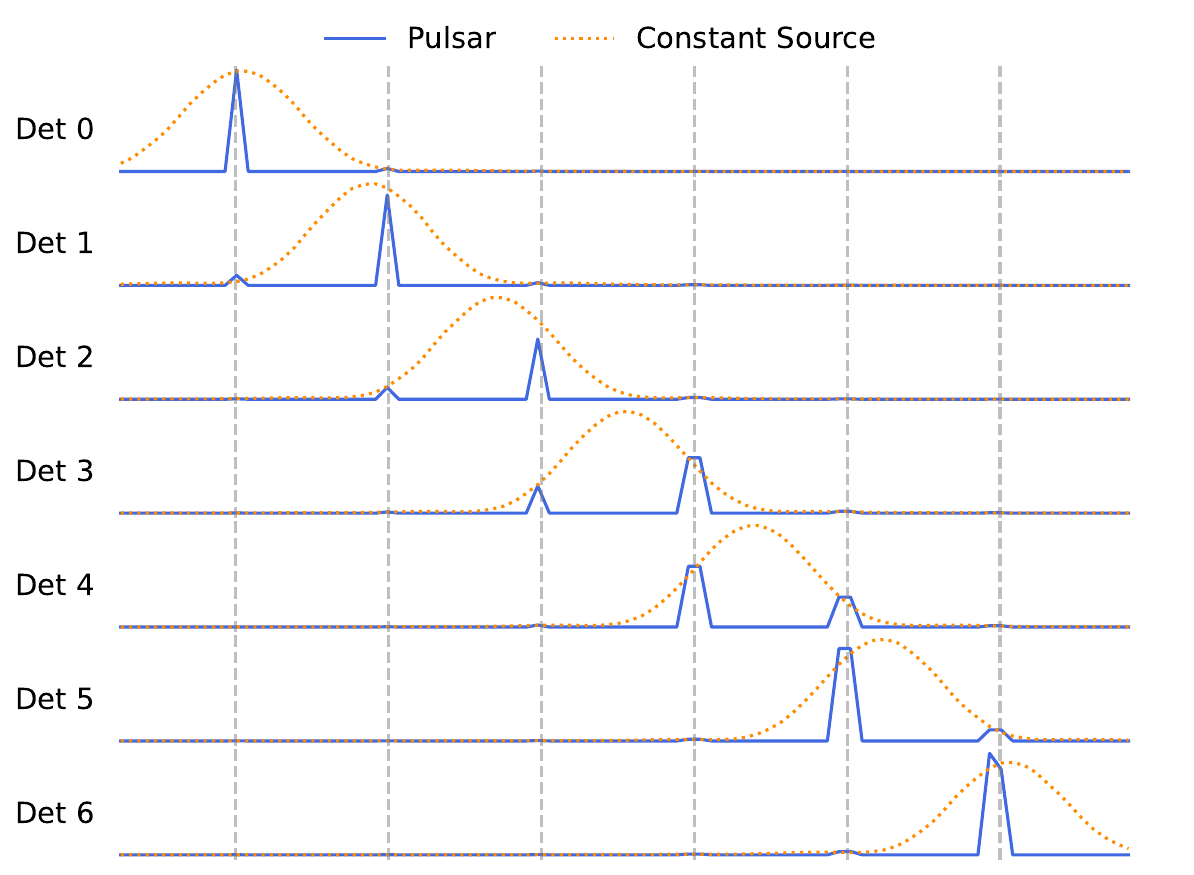}
  \caption{A schematic diagram illustrating how a simulated Crab Pulsar signal propagates across detectors aligned in the scan direction. The solid blue curves represent the simulated pulsar's periodic signal ($T \approx 33.5$~ms). The orange dotted lines show the expected signal from a constant source at the pulsar's location. Vertical dashed lines indicate the expected pulse arrival times based on the pulsar's period. This simulation uses realistic ACT parameters (scan speed and detector spacing). The apparent coincidence that successive pulses land on successive detectors is a real feature of the Crab observations, as discussed in Section~\ref{sec:methods}.}
  \label{fig:pulsar-sim}
\end{figure}

The Crab Pulsar, with its approximately 33\,ms period and well-characterized timing properties \citep{Lyne:1993:crab}, is a natural benchmark for demonstrating ACT's ability to observe known pulsars. However, detecting ms-scale periodic signals with ACT presents unique challenges. The rapid timescale precludes standard map-making techniques, necessitating direct analysis of the raw time streams. ACT's continuous scanning motion causes the pulsar signal to disperse across multiple detectors as the telescope sweeps across the source. Figure~\ref{fig:pulsar-sim} illustrates this effect, showing a simulated pulsar signal distributed across detectors along the scan direction. For a typical scan speed of 1\,deg\,s$^{-1}$ and an effective on-sky detector separation of around $2'$, the time for the source to travel between adjacent detectors is approximately 33~ms. This is, coincidentally, close to the pulsar's period ($T \approx 33.5$\,ms). Consequently, successive pulses tend to be registered by successive detectors along the direction of scan. Pulsar searches must therefore account for telescope motion and coherently combine data across detectors to maximize detection sensitivity. It is worth noting that the method we develop in this work is generic and not dependent on specific pulsar periods nor the scan strategies. It can also account for scenarios where multiple pulses are observed by a single detector as it scans across the source (i.e., for much shorter periods). The fundamental limit on the detectable pulse period is set by the instrument's sampling rate and the desired phase resolution. With a sampling rate of up to 400\,Hz (a time resolution of 2.5\,ms), our analysis can, in principle, probe periods down to the millisecond pulsar regime.

The short pulse timescale of the Crab Pulsar also demands millisecond-level timing precision over a long observation baseline to avoid pulse smearing. To address this, we utilize the ephemeris from the Jodrell Bank Pulsar Monitoring Program \cite{Lyne:1993:crab}, which has continuously monitored the Crab Pulsar since 1984.\footnote{\url{http://www.jb.man.ac.uk/~pulsar/crab.html}}

\begin{figure}[t]
  \centering
  \includegraphics[width=\linewidth]{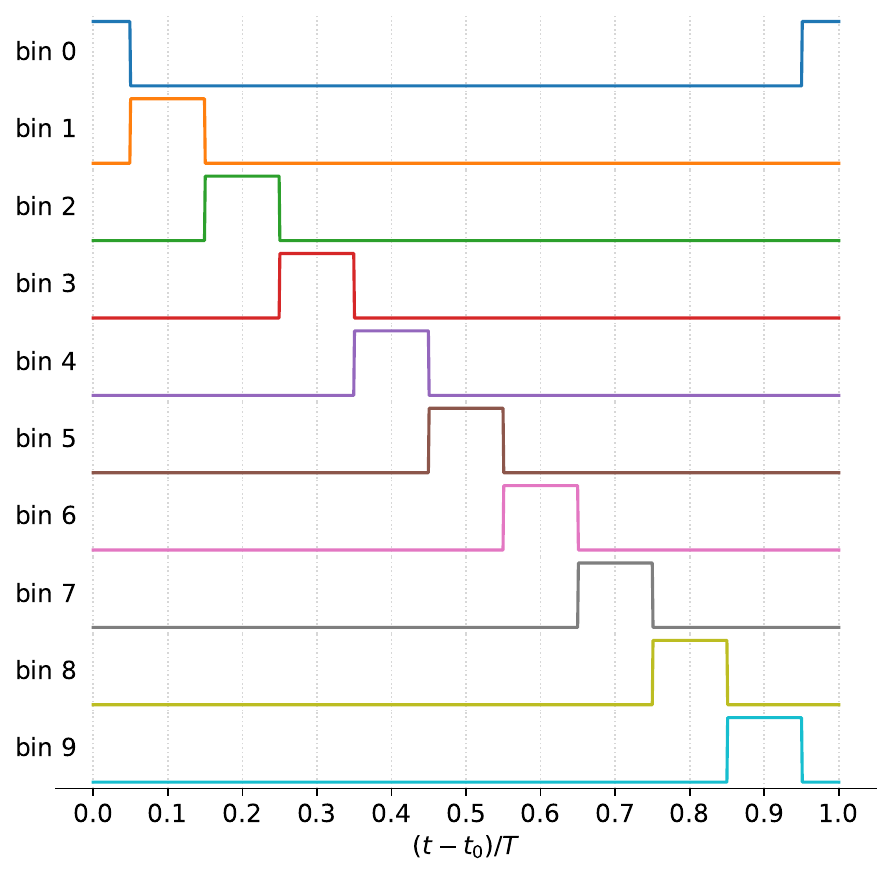}
  \caption{Boxcar phase profiles, $\eta_{i}(t)$, used in the pulsar search. Each row represents the profile for a different phase bin, showing the temporal weighting applied to each period of the data.}
  \label{fig:phase-profile}
\end{figure}

While the Crab Pulsar exhibits a complex temporal profile characterized by main and inter-pulse components \citep[e.g.,][]{Hankins:2015}, ACT's temporal resolution limits our sensitivity to these fine details. Therefore, we simply adopt a boxcar profile for our analysis, defined as:
\begin{equation}
\eta_i(t) =
\begin{cases}
1, & \text{if } \frac{i-1/2}{n_b} \leq \frac{(t - t_0) \bmod T}{T} < \frac{i+1/2}{n_b} \\
0, & \text{otherwise},
\end{cases}
\label{eq:pulsar-profile}
\end{equation}
where \(t\) is the observation time, \(t_0\) is a reference time, \(T\) is the pulsar period, \(n_b\) is the number of phase bins per period, and $i$ is the bin index (ranging from 0 to $n_{b}-1$). We choose $n_b=10$ for this analysis, resulting in bins of approximately 3.3\,ms duration. This ensures that each bin contains, on average, more than one sample for ACT's 400\,Hz sampling rate (PA5 and PA6) and about one sample for the 300\,Hz sampling rate (PA4). The resulting search profiles for the different phase bins are visualized in Figure~\ref{fig:phase-profile}. This choice of orthogonal boxcar templates ensures robust period-averaged flux measurements against profile mismatch, as we demonstrate in Appendix~\ref{app:impacts-pulse-profiles}.

\subsection{Data preprocessing}
Our data preprocessing follows the same procedures as in ACT DR6. Initial preprocessing involves deconvolving the MCE filter and the detector time constant response (see Sec.~\ref{sec:act}), before cutting malfunctioning detectors as well as short portions of the timestreams contaminated by glitches. 
We then calibrate the detector timestreams following the methodology detailed in the ACT DR6 paper \citep{ACT:Naess:2025}. First, we convert raw detector outputs within each TOD from data acquisition unit to pico-watts (pW) using bias step measurements. Next, we perform relative calibration across detectors using atmospheric loading as a flat-field signal and recalculate these flat-fields weekly to mitigate gain variations. We then calibrate our timestreams from pW to micro-Kelvin ({\textmu}K) using Uranus as an absolute calibrator. Finally, we convert these data to flux density units of millijansky (mJy), assuming effective band centers of 96, 148, and 225\,GHz for the f090, f150, and f220 bands, respectively, and accounting for the solid angle of the beam at each band.

At low temporal frequencies ($\lesssim 3$\,Hz), the dominant source of noise in the timestreams is mm emission from water vapor in Earth's atmosphere, which manifests as spatially correlated noise across detectors and creates a slowly varying signal as the telescope scans over the turbulent atmosphere. Although the atmospheric noise dominates most astrophysical signals, the Crab Pulsar is embedded within the bright Crab Nebula (with a flux density of $\sim 300$\,Jy at 96\,GHz), which is not subdominant. A simple low pass filter of the timestreams would therefore be biased by the large signal from the nebula. To avoid this, we estimate the atmospheric noise's detector-detector covariance using data from regions $>0.2^\circ$ away from the Crab Pulsar. For each sample in the time-ordered data, we then use the detector values from these $>0.2^\circ$ regions, along with the covariance matrix, to obtain a maximum-likelihood estimate of the atmospheric noise in the detectors $<0.2^\circ$ away from the pulsar. This provides a prediction of the atmospheric noise in the pulsar region that is independent of the actual pulsar and nebula signals. We create a copy of the timestream with these interpolations replacing the Pulsar nebula signal, low pass filter it with a Butterworth filter,
\begin{equation}
  \frac{1}{1 - (f/3\,\textrm{Hz})^{-10}}
\end{equation}
where $f$ is the frequency, and subtract it from the original timestream. Thus, the atmospheric noise is subtracted from the regions of the timestream containing the Crab Nebula without biasing the signal itself (at the cost of removing some signal outside this region, which we are not interested in anyways).\footnote{In practice, we also low-pass filter the noise prediction (with a cutoff at 3 Hz) before subtracting it from the data. This is because the noise prediction at frequencies above 3 Hz is dominated by uncorrelated noise, which is not useful for predicting the atmospheric noise in the target region and would only add noise to the result.}

Thermal changes in the telescope structure and other factors can affect the telescope pointing at the ${\sim}1'$ level \citep{ACT:Naess:2025}, which can significantly impact the detection significance. Thus, we fit the positions of known point sources in our TODs to provide per-TOD pointing corrections, which achieves a pointing precision of ${\sim}0.15'$, which is significantly below our beam sizes ($1'$-$2'$) as well as our pixel size ($0.5'$). However, it is worth noting that in the bright Crab Nebula region, sub-pixel pointing errors could still manifest as multiplicative noise. We account for this effect in our noise model through data splits (Section~\ref{sec:map-based-method}), which empirically captures such systematics.

\subsection{Timing corrections}
Precise timing is crucial for pulsar detection, particularly over long observation baselines. When our observation duration spans over one month, the required relative precision in the period measurement $\Delta P/P$ must be significantly better than $1.3 \times 10^{-8}$ to avoid pulse smearing. This precision requirement cannot be met with simple period models, even when accounting for period derivatives, due to both secular changes and sudden rotational ``glitches'' in the pulsar's behavior. To address this, we utilize timing solutions from the Jodrell Bank Pulsar Monitoring Program, which provides detailed ephemerides tracking these variations over decades of observations.

Converting our observed times to the pulsar's reference frame requires multiple corrections. We transform observer's Coordinated Universal Time (UTC) to Barycentric Dynamical Time (TDB, from the French \textit{Temps Dynamique Barycentrique}), which accounts for various effects including telescope-to-Earth-center light travel time, Earth-center-to-Solar-System-barycenter propagation, gravitational time dilation, and frequency-dependent dispersion delays.

We validated our timing corrections implementation against standard examples provided in the pulsar timing database. The long baseline of the timing database we validated against, spanning over 40 years, also ensures that any jitter or timing instabilities in the pulsar would have been captured and accounted for in our timing model. ACT's absolute timing accuracy of better than 1 ms and approximately 1 ms-level detector constants are sufficient for our analysis, given the Crab pulsar's 33.5 ms period.

\subsection{Map-based method}
\label{sec:map-based-method}
To construct a phase-resolved map of the Crab Pulsar, $m_{b}$, in phase bin interval $b$, we solve a modified mapmaking equation,
\begin{equation}
  d_{i}(t) = \sum_{b} \eta_{b}(t) \left[P_{i}(t) m_b + n_{i}(t)\right],
\end{equation}
where $d_i(t)$ is the TOD, $i$ is the detector index, and $t$ is time. $\eta_{b}(t)$ represents the pulsar timing profile for phase bin $b$, as defined in Equation~\eqref{eq:pulsar-profile}. $P_{i}(t)$ is the pointing matrix, mapping the sky to detector $i$ at time $t$ based on the telescope's motion. $n_i(t)$ is the noise per detector as a function of time.

By leveraging the properties $\eta_{i}(t)\eta_{j}(t)=0$ when $i\neq j$ (as seen in Figure~\ref{fig:phase-profile}), we derive the maximum-likelihood solution to the mapmaking equation:
\begin{equation}
  \label{eq:ml-solution}
  m_{b} = (P^{T}\eta_b^T N^{-1}\eta_b P)^{-1}P^{T}\eta_b^T N^{-1} d.
\end{equation}
This is equivalent to making a separate map for the samples that fall inside each of the box-car profiles, $\eta_{b}$. $N$ is the noise covariance matrix across detectors and the samples in phase-bin $b$, which we assume to be stationary in time and diagonal across detector space. This assumption holds for our high-pass filtered data ($\gtrsim 3\,$Hz), where the detector noise is dominated by uncorrelated components and the pulsar signal resides (at ${\sim} 30\,$Hz).

\begin{figure}[t]
    \centering
    \includegraphics[width=\linewidth]{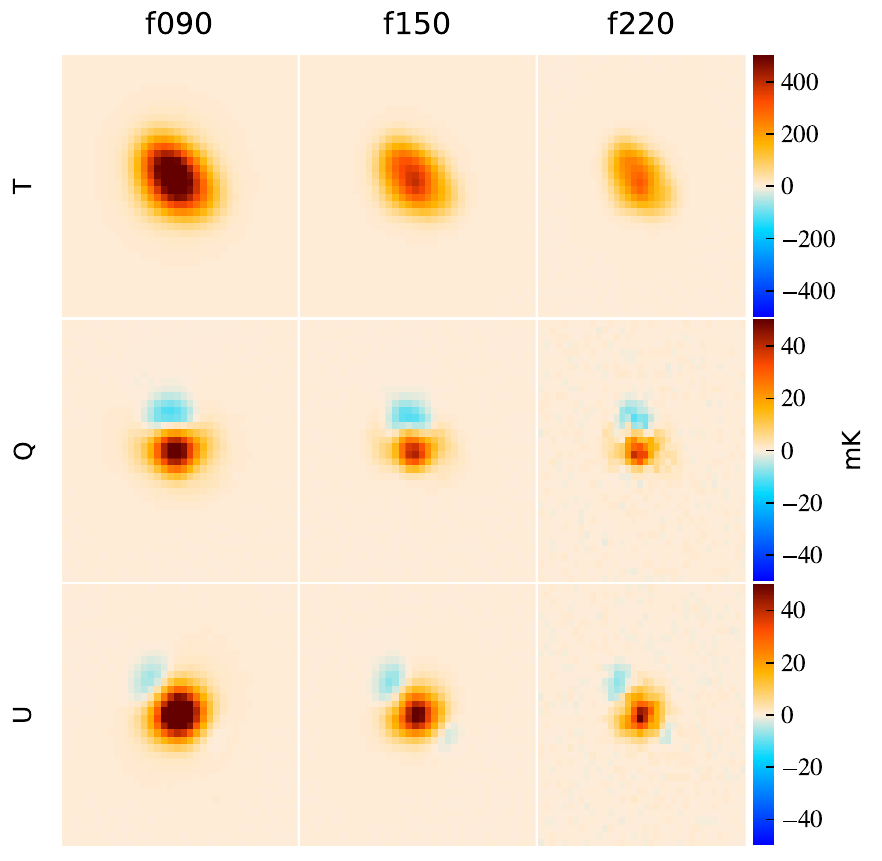}
    \caption{Maps of the Crab Nebula from phase bin $b = 0$, showing a 0.4$^\circ$ region centered on the Crab Pulsar position. Rows display total intensity (T), Q, and U polarization maps of the nebula, where the pulsar itself is invisible. Columns illustrate the frequency dependence of the nebular emission.}
    \label{fig:per_freq_tqu_split0}
\end{figure}

\begin{figure*}[t]
  \centering
  \includegraphics[width=\linewidth]{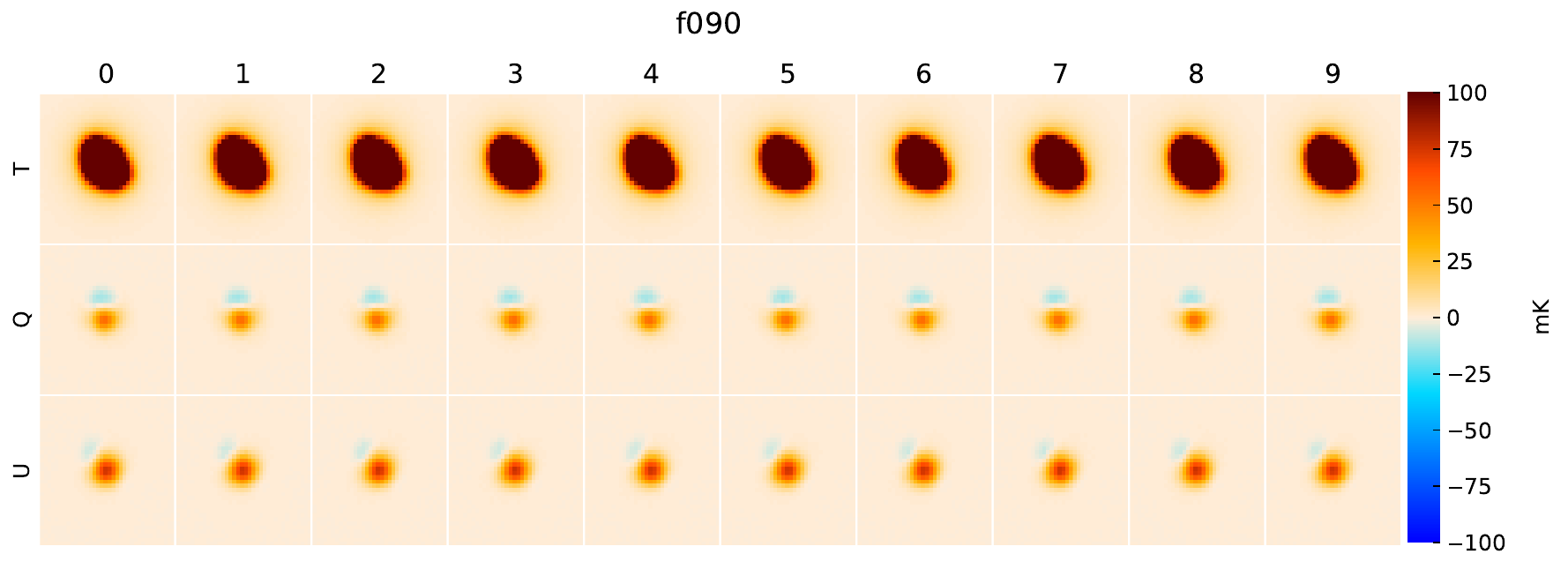}
  \caption{Maps of the Crab Nebula for the f090 band across all 10 phase bins. Each column corresponds to a phase bin (0 to 9). Rows show total intensity (T), Q, and U polarization. The maps are visually identical across phase bins, indicating that the dominant emission from the Crab Nebula is constant in time and overwhelms any pulsar signal.}
  \label{fig:phase_resolved_maps_f090}
\end{figure*}

The mapmaking process yields 10 phase-resolved maps, one for each phase bin, constructed by stacking the 213 (f090), 326 (f150), and 113 (f220) individual TODs. Figure~\ref{fig:per_freq_tqu_split0} shows these maps for phase bin $b = 0$ at different frequencies centered around the expected location of the Crab Pulsar. Maps for other phase bins are visually indistinguishable from this one. As an example, Figure~\ref{fig:phase_resolved_maps_f090} shows the maps for the f090 band across the phase bins. The similarity of maps across phase bins is due to the dominant, constant emission from the Crab Nebula, which overwhelms any potential pulsar signal in the individual phase-resolved maps. The pulsar signal, if detectable, would manifest as subtle differences between the maps across the phase bins.

\begin{figure*}[t]
    \centering
    \includegraphics[width=1\linewidth]{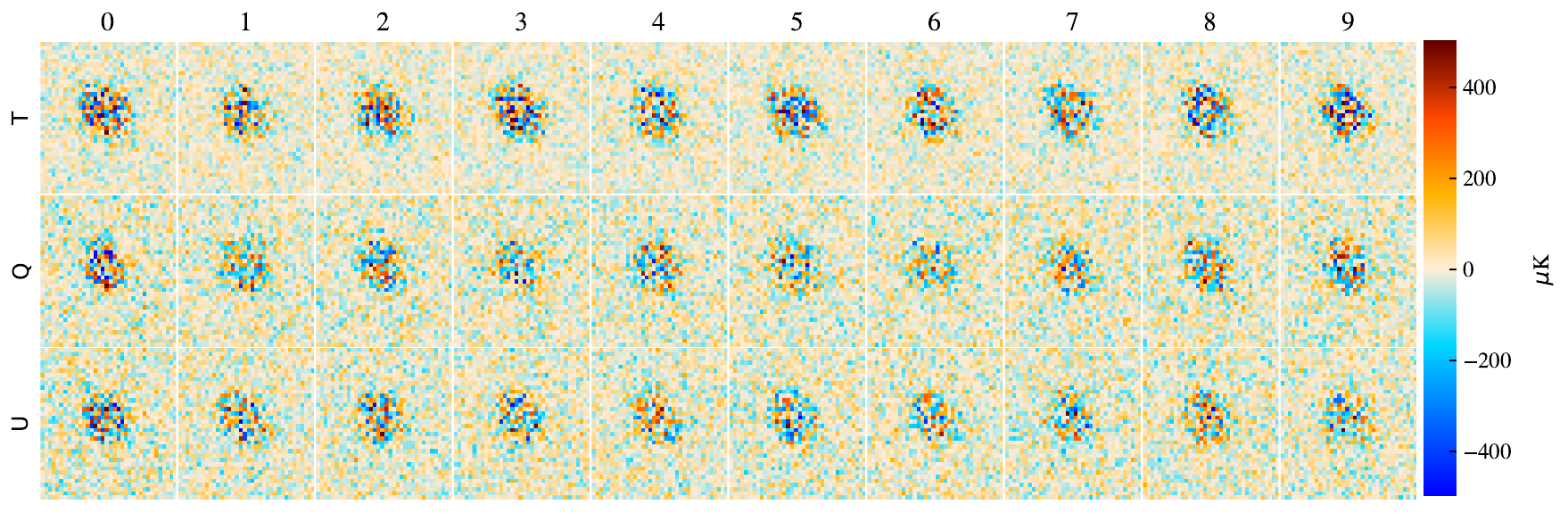}
    \caption{Difference maps at f090 across phase bins, obtained by subtracting the mean map (averaged across all phase bins) from each individual phase-resolved map. Rows show T, Q, and U maps, and columns represent different phase bins. The enhanced noise level in regions of high emission indicates multiplicative noise, potentially caused by gain miscalibration. Similar effects are observed at the other two frequencies.}
    \label{fig:diff_map_f090}
\end{figure*}

To isolate the pulsar signal, we compute the mean map across all phase bins and subtract it from each individual phase-resolved map, effectively removing the Crab Nebula's contribution. We denote the resulting difference maps as $\delta m_{b}\equiv m_{b} - \langle m_{b}\rangle$. Figure~\ref{fig:diff_map_f090} shows the resulting difference maps for the f090 band. These difference maps reveal that the noise level is approximately modulated by the Crab Nebula's emission profile -- a characteristic signature of multiplicative noise. Such noise can arise from coupling between the bright sky signal and systematic errors in, for example, gain calibration or telescope pointing. For instance, a gain calibration error causes individual detectors to disagree on the amplitude of the sky signal; these discrepancies manifest as apparent noise that is proportional to the underlying sky signal. In ACT DR6, the gain calibrations are known to approximately $1\%$ precision, which is consistent with the level of residual seen in Figure~\ref{fig:diff_map_f090}, and is thus likely the source of the observed multiplicative noise. If unaccounted for, such multiplicative noise could lead to under-estimation of the uncertainties of our flux measurements. To mitigate this, we divide our TODs into four groups of independent data, or `splits', and repeat the mapmaking and subtraction process for each split. This allows us to develop a more robust noise model numerically based on the differences between splits, each of which has independent data, thereby accounting for the multiplicative noise contribution.

To estimate the flux of the Crab Pulsar, we apply a matched filter to the phase-resolved difference map, $m$,
\begin{equation}
  \label{eq:match-filter}
  F = \frac{B^T M^{-1}\delta m}{\text{diag}(B^T M^{-1}B)} \equiv \frac{\rho}{\kappa}
\end{equation}
where $B$ is the beam covariance matrix, $M$ is the pixel-pixel noise covariance estimated from the independent data splits, and $\delta m$ is a phase-resolved difference map. Each pixel in the resulting match-filtered map, $F$, represents the maximum-likelihood estimate of the flux if a point source were centered at that pixel. Here we have defined $\rho\equiv B^T M^{-1}m$ to represent the unnormalized beam response from $m$, with $\kappa \equiv \text{diag}(B^T M^{-1}B)$ being the normalization factor. The flux estimate is then given by $\rho/\kappa$, with an uncertainty of $\kappa^{-1/2}$. Since the maps are centered on the Crab Pulsar's known location, we extract the flux estimates at the central pixel of $F$. As the pulsar's phase profile at millimeter wavelength is poorly constrained, we search across all phase bins and derive the period-averaged flux density. This approach is robust to uncertainties in the pulse profile shape, as demonstrated in Appendix~\ref{app:impacts-pulse-profiles}.

\subsection{Signal-injection tests}
\label{sec:signal-injection-test}

\begin{table}[t]
\centering
\begin{tabular}{c c c}
\hline
$b$ & $\Delta$RA ['] & $\Delta$Dec ['] \\
\hline
0 & 0 & 0 \\
1 & +6 & +6 \\
2 & 0 & +6 \\
3 & -6 & +6 \\
4 & -6 & 0 \\
5 & -6 & -6 \\
6 & 0 & -6 \\
7 & +6 & -6 \\
8 & +6 & 0 \\
\hline
\end{tabular}
\caption{Simulated source locations in arcminutes relative to the Crab Pulsar, and the corresponding phase bin ($b$) in which each source was injected.}
\label{tab:simulated_sources}
\end{table}

\begin{figure}[t]
  \centering
  \includegraphics[width=0.7\linewidth]{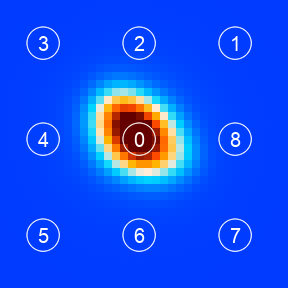}
  \caption{Locations of simulated pulsar signals injected into the f090 data, overlaid on a map of the Crab Nebula region. Circles represent the ACT beam size, with the number inside indicating the phase bin ($b$) of the injected signal (see Table~\ref{tab:simulated_sources} for precise coordinates). One pulsar is co-located with the Crab Nebula.}
  \label{fig:injection-location}
\end{figure}

To validate the pipeline described above, we performed signal-injection tests. We injected nine simulated, unpolarized pulsars into the f090 TODs used for the Crab Pulsar search, ensuring that the simulated signals exhibited consistent noise properties with the real search. Each injected pulsar had a fixed peak flux density of 50\,mJy and a boxcar profile, as defined in Equation~\eqref{eq:pulsar-profile}, with each pulsar's signal appearing in a distinct phase bin at the same pulse period as the Crab Pulsar. The pulsars were arranged on a $3\times 3$ grid with $6'$ spacing between adjacent sources, with one pulsar located at the same coordinates as the Crab Pulsar. Figure~\ref{fig:injection-location} shows the locations and phase bin of each injected pulsar signal relative to the Crab Nebula region. This grid arrangement allowed us to test our pipeline with different noise realizations while using the same set of f090 TODs, ensuring that the results are representative of the statistical uncertainties in the real data. Table~\ref{tab:simulated_sources} summarizes the location (relative to the Crab Pulsar) and the expected phase bin for each injected pulsar signal.

\begin{figure*}[t]
\centering
\includegraphics[width=\linewidth]{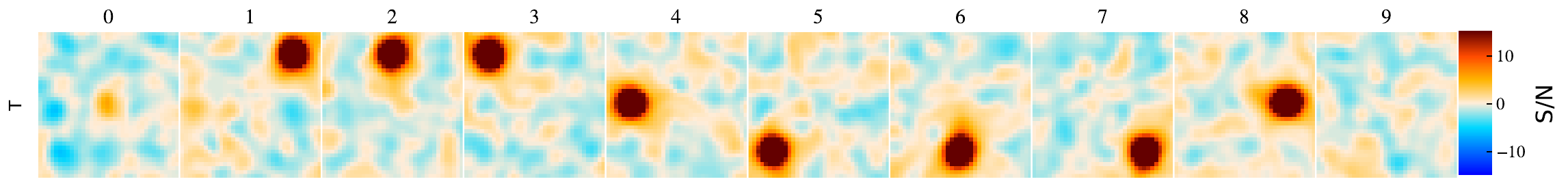}
\caption{Signal-to-noise maps (with cross-phase-bin mean subtracted) for the simulated f090 data with injected sources in nine phase bins (see Table~\ref{tab:simulated_sources}). Each panel shows the S/N map for a single phase bin centered on the Crab Pulsar, with noise estimated using data splits. Phase bin 9 (last panel) serves as a control with no injected source. Phase bin 0 (first panel) has the simulated pulsar on top of the Crab Nebula, resulting in lower signal-to-noise due to higher multiplicative noise in that region.}
\label{fig:sim_snr}
\end{figure*}

Figure~\ref{fig:sim_snr} shows the resulting signal-to-noise maps from our phase-resolved mapmaking pipeline. The pipeline successfully recovers each injected pulsar signal, with a clear peak in the S/N map at the predicted location and phase bin. The signal-to-noise ratio for the pulsar injected at $b = 0$ (co-located with the Crab Pulsar) is significantly lower than for the other injected sources. This is due to the effects of multiplicative noise acting on the pulsar nebula, as discussed in Section \ref{sec:map-based-method}.

\begin{table}[t]
  \centering
  \begin{tabular}{c | c c c | c c c}
    \hline
     & \multicolumn{3}{c|}{\textbf{map-fit}} & \multicolumn{3}{c}{\textbf{tod-fit}} \\
    $b$ & I [mJy] & $\sigma$ [mJy] & SNR & I [mJy] & $\sigma$ [mJy] & SNR \\
    \hline
    0 & 41.2 & 7.6  & 5.4  & 56 & 21 & 2.7  \\
    1 & 47.8 & 1.1  & 44 & 46.7 & 1.1  & 44 \\
    2 & 51.2 & 1.2  & 44 & 49.1 & 2.1  & 24 \\
    3 & 50.7 & 1.0  & 50 & 50.2 & 1.2  & 41 \\
    4 & 52.9 & 1.2  & 46 & 51.9 & 1.8  & 28 \\
    5 & 47.9 & 1.0  & 46 & 46.5 & 1.3  & 37 \\
    6 & 50.2 & 1.1  & 45 & 49.2 & 1.6  & 32 \\
    7 & 49.4 & 1.1  & 44 & 48.7 & 1.5  & 33 \\
    8 & 50.1 & 1.1  & 44 & 47.6 & 2.4  & 20 \\
    \hline
  \end{tabular}
  \caption{Comparison of map-level per-pixel fit and TOD-level fit for the injected pulsar signals results at each phase bin ($b$), where $I$ and $\sigma$ represent the flux and its uncertainty (standard deviation) in total intensity in units of mJy, respectively. SNR represents the signal-to-noise ratio, defined as $I / \sigma$.}
  \label{tab:comparison}
\end{table}

The left-hand side of Table~\ref{tab:comparison} (labeled ``map-fit'') presents the flux measurements for each injected pulsar, extracted from each matched-filtered map at its corresponding phase bin. For the majority of injected sources, the recovered flux is within $\pm 1\sigma$ of the injected flux (50 mJy), detected at $\gtrsim 40 \sigma$ significance. A few sources exhibit mild deviations of approximately $2\sigma$, potentially indicating unaccounted-for correlated noise between sources. This correlated component is expected as all sources were injected into the same TOD. We therefore conclude that our search pipeline accurately recovers injected pulsar signals at their predicted phase bins, with a flux uncertainty ranging from approximately 1\,mJy to 7\,mJy, depending on whether the source is superimposed on the Crab Nebula.

\subsection{Alternative method: TOD-based matched-filtering}
To further validate our search pipeline, we developed an alternative search algorithm that directly fits for the pulsar signal in the TOD. We model the TOD, $d_{i}(t)$, as
\begin{equation}
  \label{eq:tod-fit}
  d_{i}(t) = A T_{i}(t) + n_{i}(t),
\end{equation}
where $A$ is the flux to be estimated, and $T_{i}(t)$ is a time-dependent signal template with a unit flux, defined as
\begin{equation}
  T_{i}(t) = P_{i}(t) \eta(t) B s.
\end{equation}
Here, $P_{i}(t)$ is the pointing matrix, $\eta(t)$ is the temporal profile of a point source, which may or may not be periodic, $B$ is the beam covariance matrix, and $s$ is a map that is zero everywhere except at the location of the source, where it has a value of one. Thus, $Bs$ represents a map with a unit-flux point source. Estimating the flux of a time-varying signal therefore requires solving for $A$ in Equation~\eqref{eq:tod-fit}. The maximum likelihood solution for $A$ is
\begin{equation}
  \label{eq:tod-fit-ml-solution}
  \hat{A} = (T^{T} N^{-1} T)^{-1} T^{T} N^{-1} d,
\end{equation}
where $N$ is the noise covariance matrix, as in Equation~\eqref{eq:ml-solution}. Equation~\eqref{eq:tod-fit-ml-solution} provides a general method for measuring any time-varying signal from a point source in the timestream, given knowledge of its temporal profile, $\eta(t)$. In the case of a pulsar search, we set $\eta(t)$ to be a periodic signal based on the Crab Pulsar timing model and a boxcar pulse profile. We fit a set of 10 temporal profiles, each corresponding to one of the 10 phase bins.

We preprocess the TODs in the same way as used in the phase-resolved mapmaking method. To account for multiplicative gain, we divide the TODs into 20 independent splits. For each split, we estimate the fluxes across the 10 phase bins and subtract the median flux across phase bins to remove the contribution from the Crab Nebula, which is expected to be present in all phase bins. We repeat this process for all 20 splits and estimate the mean flux and its corresponding uncertainty for each phase bin using the 20 independent flux estimates.

\begin{figure}[t]
\centering
\includegraphics[width=\linewidth]{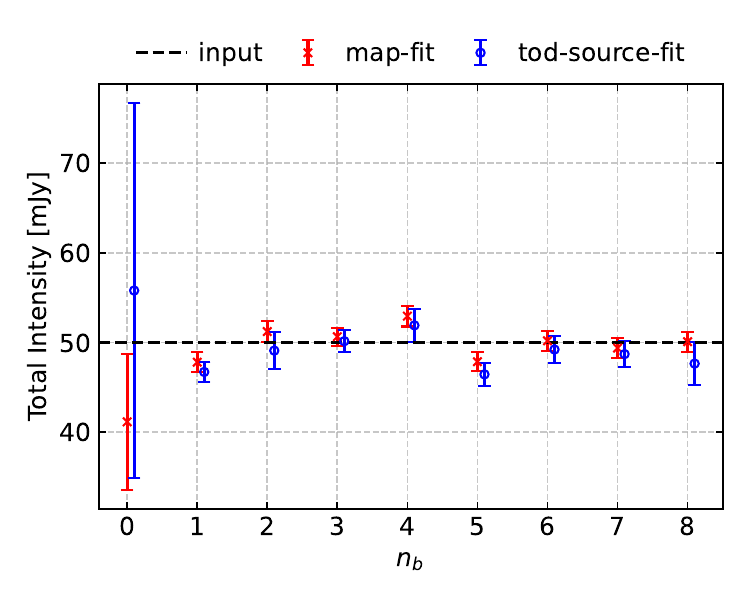}
\caption{Comparison of the reconstructed pulsar fluxes at each phase bin from using our simulated maps between the two approaches. Dashed line represents the flux density (50 mJy) of the injected signal. $n_{b}$ represents phase bin number. Data points shifted slightly for visualization purpose.}
\label{fig:sim-comparison}
\end{figure}

\begin{figure*}[t]
\centering
\includegraphics[width=\linewidth]{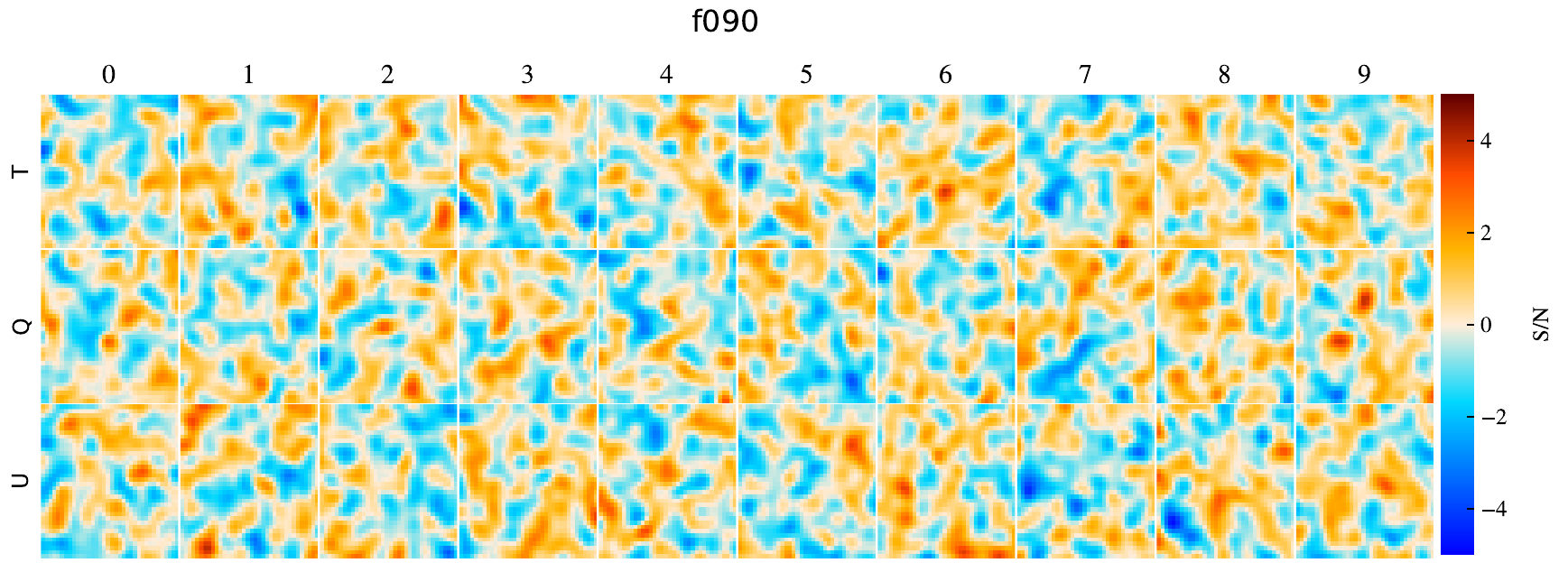}
\includegraphics[width=\linewidth]{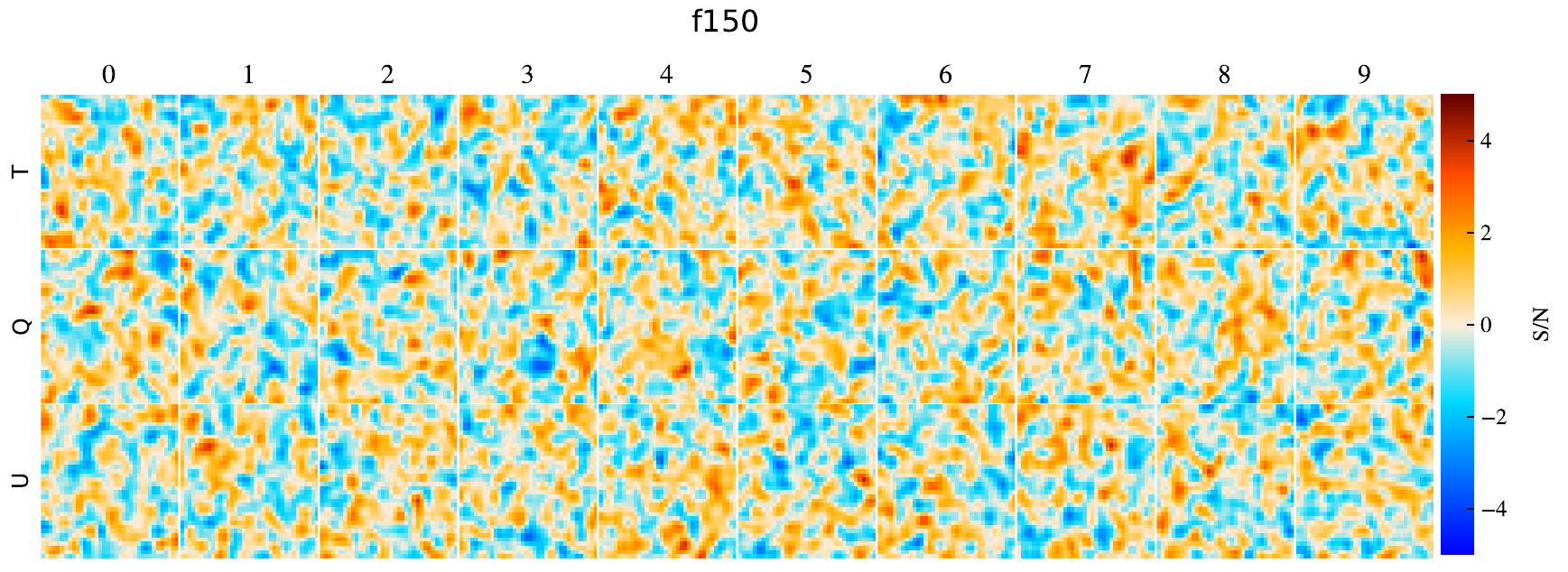}
\includegraphics[width=\linewidth]{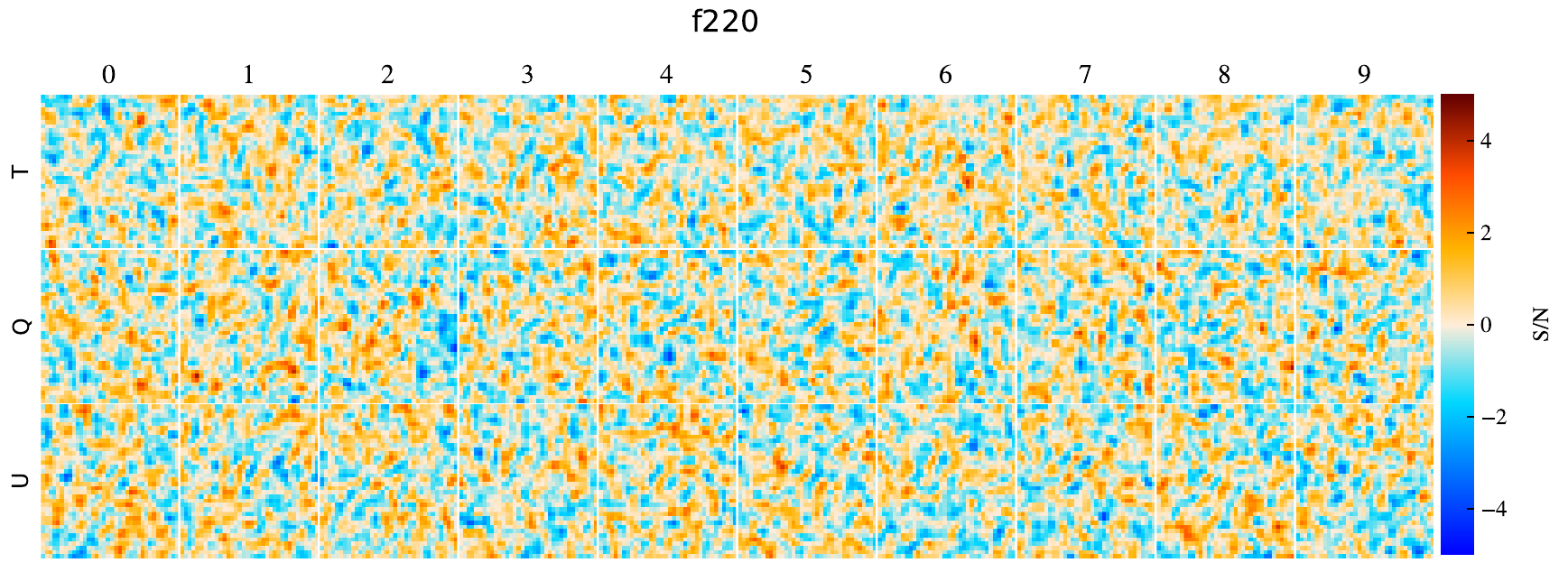}
\caption{Signal-to-noise maps for all phase bins across frequencies with cross-phase-bin mean maps subtracted. The Crab Pulsar signal is centered in each map. No single bin shows significant detection.}
\label{fig:snr_map}
\end{figure*}

To validate this pipeline, we applied it to the same signal-injection tests described in Section~\ref{sec:signal-injection-test}. We compare the estimated flux for each injected pulsar in its predicted phase bin with the results from the mapmaking approach in Figure~\ref{fig:sim-comparison} (also listed in the right-hand side of Table~\ref{tab:comparison} labeled ``tod-fit''). The comparison shows that this TOD-based fit correctly reconstructs the injected pulsar signals with a flux of 50 mJy at a significance level of $\gtrsim 20\sigma$ for the majority of the phase bins, and demonstrates consistency with our baseline mapmaking results. One notable difference is in the larger uncertainties, which is most evident in the $b=0$ phase bin. This was initially surprising. We later understood it to be a consequence of modeling error: Equation~\eqref{eq:tod-fit} models the TOD as a combination of a point source and noise. However, the Crab Nebula's extended emission ($\sim 5'$), violates this point-source assumption, causing its flux to be misinterpreted as noise and inflating the measurement uncertainty. This effect is strongest when the pulsar is co-located with the Crab Nebula, resulting in a factor of 3 increase in the uncertainty for the 0th phase bin. The mapmaking method, by jointly fitting across multiple pixels, better captures the extended emission from the nebula. We extended our TOD-based model to include more degrees of freedom by jointly fitting multiple sources to account for the extended nebula. Preliminary results were consistent with the mapmaking approach. This confirms that the discrepancy is due to the limited degrees of freedom in the single-source TOD model, but also indicates that a TOD-based approach remains viable.

\section{Results and Discussion}
\label{sec:results}

\begin{figure}[t]
\centering
\includegraphics[width=\linewidth]{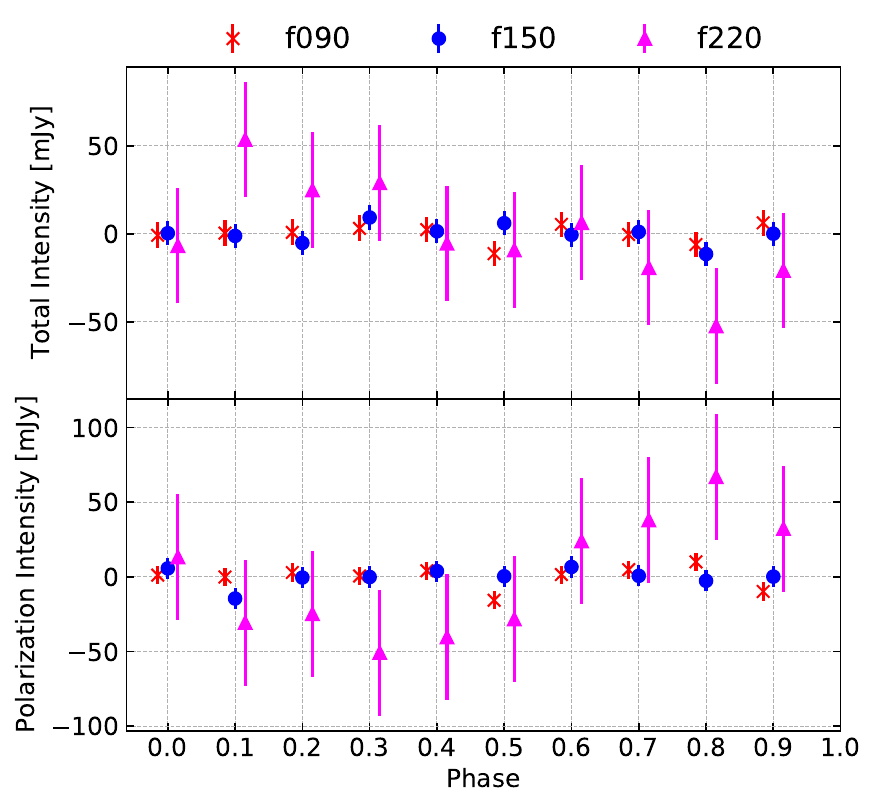}
\caption{Map-based pulsar flux measurements. Fluxes from different frequencies are shifted for visualization only.}
\label{fig:flux-measurements}
\end{figure}

Having validated the phase-resolved mapmaking method with signal injection tests and cross-checked the results with a TOD-based fitting algorithm, we present the results from our phase-resolved mapmaking pipeline. Figure~\ref{fig:snr_map} shows the signal-to-noise map, defined as $\rho\kappa^{-1/2}$, measured from each frequency band and phase bin in temperature and polarization (T/Q/U). No phase bins show significant detection of any flux at the expected location in the center.

Figure~\ref{fig:flux-measurements} shows the pulsar flux measurements at different frequency bands and phase bins. The flux measurements are consistent with zero across all phases. This allows us to constrain the period-averaged flux density of the pulsar to be 4.6\,mJy / 4.4\,mJy / 20.7\,mJy at 95\% confidence at the f090 / f150 / f220 bands in total intensity, $I$, and 3.8\,mJy / 4.5\,mJy / 26.6\,mJy at 95\% confidence in total polarization, $P = \sqrt{Q^{2} + U^{2}}$. While one might expect the sensitivity in polarization to be $\sqrt{2}$ times worse than in total intensity due to photon noise statistics, in our case the uncertainties are dominated by systematic effects, primarily multiplicative gain errors. These systematics impact the total intensity measurement more significantly than polarization, as the Crab Nebula is only weakly polarized at $\sim 7$\% in the mm-band \citep{Aumont:2010, Page:2007}.

\begin{figure*}[t]
  \centering
  \includegraphics[width=0.8\linewidth]{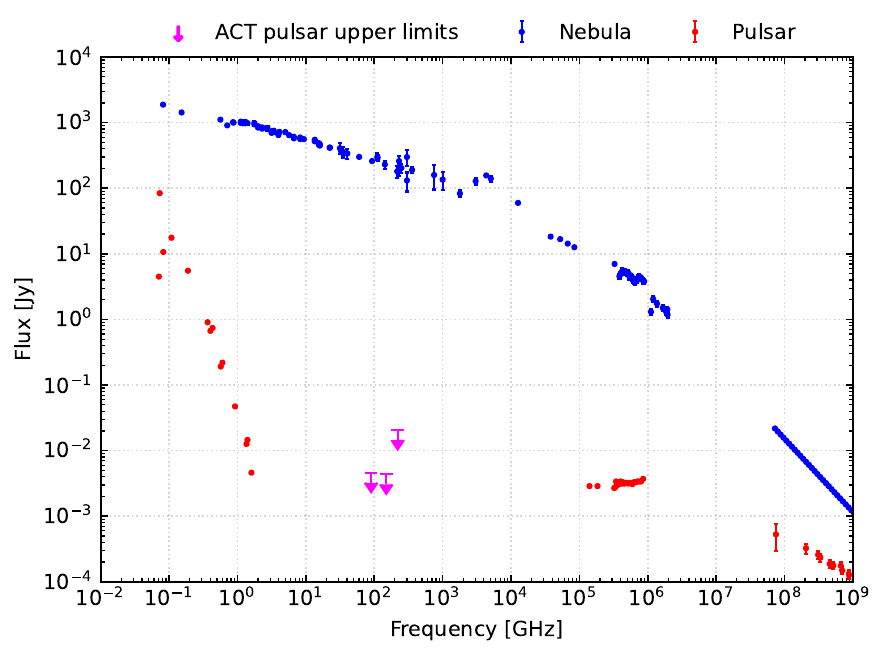}
  \caption{Spectral energy distribution (SED) of the average emission of the Crab nebula (blue) and the phase-averaged emission of the Crab pulsar (red). Data plotted are taken from \citep{Buhler:2014} which reproduced measurements from \cite{Meyer:2010,Kuiper:2001,Sollerman:2000,Tziamtzis:2009}. ACT's 95\% upper limits in the Crab Pulsar's phase-averaged flux density are shown in magenta.}
  \label{fig:crab-spectra}
\end{figure*}

The Crab Pulsar, powered by a young, rapidly rotating neutron star \citep[see, e.g.,][]{Gold:1968}, exhibits broadband emission spanning radio to gamma-ray energies. This emission arises primarily from synchrotron radiation at lower energies and inverse Compton scattering at higher energies \citep{Blandford:2002, Buhler:2014}, with particularly strong emission observed up to 400 GeV in gamma rays \citep{Buhler:2014}. Figure~\ref{fig:crab-spectra} shows the low-energy part of the phase-averaged spectral energy distribution (SED) of the period-averaged Crab Pulsar emission from the literature compared to the Crab Nebula, which reveals a notable gap in our understanding of the pulsar's spectral energy distribution in the millimeter and sub-millimeter range, particularly within ACT's frequency bands, where it remains unclear whether the spectrum flattens or begins to rise between radio and infrared wavelengths. Our period-averaged flux density upper limits provide valuable constraints on the pulsar's integrated emission at these frequencies, suggesting that a significant flattening or upturn in its spectrum is unlikely. These results provide empirical constraints relevant to models invoking a transition from coherent to incoherent emission at higher frequencies \citep{Michel:1978}, where a spectral upturn might be expected.

It is important to consider potential systematic uncertainties that could affect our flux density measurements. Because we are measuring the period-averaged flux density, we are not sensitive to uncertainties in the precise shape of the pulsar's pulse profile. However, other systematic effects, such as pointing errors and mis-calibration, can manifest as multiplicative noise. We mitigated these effects by numerically estimating the noise covariance matrix using independent data splits, as discussed in Section~\ref{sec:map-based-method}.

To quantify the impact of these systematics, we can compare the flux uncertainties in regions with and without significant multiplicative noise. In our signal injection tests (Section \ref{sec:signal-injection-test}), where we injected simulated pulsars into regions away from the bright Crab Nebula, we achieved flux uncertainties of approximately 1\,mJy per phase bin at f090. This represents the expected level of statistical uncertainty in our dataset. In contrast, the flux uncertainties when the pulsar is injected in the Crab Nebula region are significantly larger, approximately 7\,mJy per phase bin at f090, due to multiplicative noise. This suggests that our upper limits are dominated by systematic uncertainty, which makes up 97\% of our noise budget.

The choice of analysis method also has significant impact on our sensitivity to the pulsar signal. As discussed in Section~\ref{sec:methods}, we initially explored both a phase-resolved mapmaking approach and a TOD-based fitting method. While both methods yielded consistent flux measurements on the injected pulsar signal, the mapmaking approach proved more robust in the presence of the extended Crab Nebula. The TOD-based method, which models the sky as a point source plus noise, struggled to accurately account for the nebula's emission, leading to inflated uncertainty estimates. Furthermore, the mapmaking approach offered crucial advantages for diagnosing systematic effects. For example, we observed that the noise level in our phase-resolved difference maps was modulated by the Crab Nebula's emission profile (as illustrated in Figure~\ref{fig:diff_map_f090}), indicating the presence of multiplicative noise. This effect would have been considerably harder to identify using only the TOD. Therefore, we recommend the mapmaking approach as the preferred method for future searches for faint, periodic signals in a complex background in CMB experiments. Its robustness to extended emission outweigh the TOD-based method's simplicity in this context. Furthermore, the mapmaking approach allows for easier visual inspection of the data at various stages of processing, making it more straightforward to diagnose systematic errors such as multiplicative noise, which would be difficult to identify directly in the TOD. The TOD-based method, with its simplicity and few degrees of freedom, may still be useful in cases with highly correlated noise where full covariance modelling between detectors is necessary, though a full exploration of this is beyond the scope of this work.

\section{Conclusion}
\label{sec:conclusion}
In this paper, we have demonstrated a novel method for probing sub-second periodic transients, such as millisecond pulsars with known periods, in millimeter-band data from high-precision ground-based CMB experiments like ACT, achieving mJy-level flux measurements without targeted observations. 

While other instruments have observed pulsars at millimeter wavelength \citep{Mignani:2017,Camilo:2007,Torne:2020,Torne:2022}, ACT provides a unique combination of high cadence, broad sky coverage, and close-to-simultaneous multi-frequency observations. This allows us not only to search for these transient events but also to constrain their spectral properties with unprecedented precision, filling a critical gap in our understanding of pulsar emission mechanisms in the millimeter regime where spectral measurements are scarce. We accomplished this primarily through a phase-resolved mapmaking approach, which leverages the pulsar's known periodicity to isolate its signal. We also developed a template-based timestream fit, which provided an independent validation of our results, although its sensitivity was limited by the extended emission from the Crab Nebula.

It is worth noting that we obtained these constraints on the Crab Pulsar's millimeter emission despite the fact that ACT's scan strategy was optimized for CMB mapping rather than pulsar studies. In the f090 band, for instance, only 3.6 hours out of the total 28.6 hours of Crab-containing observations were actually spent on the pulsar itself (representing around $\approx 13\%$ of the samples). A dedicated pulsar observation, in which the telescope continuously stares at the pulsar, making the Crab Nebula a constant background and eliminating time spent off-source, would deliver roughly eight times more effective integration time for the observation span. This highlights the untapped potential of CMB datasets for time-domain astrophysics.

Although we did not detect the Crab Pulsar's millimeter emission, our derived 95\% confidence upper limits of 4.6\,mJy, 4.4\,mJy, and 20.7\,mJy at 96\,GHz, 148\,GHz, and 225\,GHz, respectively, provide constraints on its period-averaged flux density, disfavoring a significant flattening or inversion of its spectrum at these wavelengths. Furthermore, the challenges we addressed in this work, including mitigating multiplicative noise from the bright Crab Nebula and developing robust noise models, will inform future searches for fast signals in CMB survey data. The algorithms developed here can be readily extended to measure millimeter flux from other radio pulsars and magnetars, a direction we will explore in future work. The upcoming Simons Observatory (SO), with its faster mapping speed and coverage of over 40\% of the sky every 1--2 days, is expected to increase the detection sensitivity to millisecond transients by a factor of three. This work establishes the viability of CMB experiments as powerful transient probes in the millimeter regime and serves as a crucial pathfinder, paving the way for future searches of fast millimeter transients using next-generation facilities like SO and CCAT.

\vspace{1em}
\noindent YG acknowledges supports from the University of Toronto's Eric and Wendy Schmidt AI in Science Postdoctoral Fellowship, a program of Schmidt Sciences. J.P.H., the George A. and Margaret M. Downsbrough Professor of Astrophysics, acknowledges the Downsbrough heirs and the estate of George Atha Downsbrough for their support. Support for ACT was through the U.S.~National Science Foundation through awards AST-0408698, AST-0965625, and AST-1440226 for the ACT project, as well as awards PHY-0355328, PHY-0855887 and PHY-1214379. Funding was also provided by Princeton University, the University of Pennsylvania, and a Canada Foundation for Innovation (CFI) award to UBC. ACT operated in the Parque Astron\'omico Atacama in northern Chile under the auspices of the Agencia Nacional de Investigaci\'on y Desarrollo (ANID). The development of multichroic detectors and lenses was supported by NASA grants NNX13AE56G and NNX14AB58G. Detector research at NIST was supported by the NIST Innovations in Measurement Science program. Computing for ACT was performed using the Princeton Research Computing resources at Princeton University, the National Energy Research Scientific Computing Center (NERSC), and the Niagara supercomputer at the SciNet HPC Consortium. SciNet is funded by the CFI under the auspices of Compute Canada, the Government of Ontario, the Ontario Research Fund–Research Excellence, and the University of Toronto. We thank the Republic of Chile for hosting ACT in the northern Atacama, and the local indigenous Licanantay communities whom we follow in observing and learning from the night sky. The Dunlap Institute is funded through an endowment established by the David Dunlap family and the University of Toronto.

\appendix

\section{ACT Detector Readout}\label{app:downsampling}

The detector readout by the MCE has been described in \citet{ACT:Battistelli:2008,ACT:Battistelli:2008b,ACT:Swetz:2011,Thesis:Hasselfield:2013}, but since the relevant details to derive the low pass filter 3\,dB cutoff and recorded data rate for PA4--6 are scattered through other sources \citep{ACT:Henderson:2016,ACT:Koopman:2018} or are not publicly documented, we provide a summary here.

The MCE was a time-domain multiplexed readout system, containing up to 32 read out lines. Each readout line was connected to $N$ detectors, which it read out sequentially before returning to the first detector and repeating. The multiplex switching time was 500\,kHZ such that the sampling rate of a whole readout line was $f_{\mathrm{raw}} = 500\,\mathrm{kHz} / N$. The data stream for each detector was then passed through with a four pole, low pass digital filter with the following transfer function:\footnote{\url{https://e-mode.phas.ubc.ca/mcewiki/index.php/Digital_4-pole_Butterworth_Low-pass_filter}}

\begin{equation}
  H(z) = 2^{-g}\left( \frac{1 + 2\,z^{-1} + z^{-2}}
                     {1 + b_{1,1}\,z^{-1} + b_{1,2}\,z^{-2}} \right)
         \left( \frac{1 + 2\,z^{-1} + z^{-2}}
                     {1 + b_{2,1}\,z^{-1} + b_{2,2}\,z^{-2}} \right).
  \label{eq_mce_transfer_function}
\end{equation}

\noindent where $z = \exp(i\,2\pi\,f/f_{\mathrm{raw}})$. For given $b_{x,y}$ coefficients, the filter cutoff was therefore dependent on $N$ via $f_{\mathrm{raw}}$.

The filter coefficients used for AdvACT starting in 2017 were $b_{1,1} = -31432\times2^{-14}$, $b_{1,2} = -29904\times2^{-14}$, $b_{2,1} = 15200\times2^{-14}$, $b_{2,2} = 13664\times2^{-14}$ and $g = 2^8$.  After being filtered, only one out of every $M^{\mathrm{th}}$ samples was recorded, thereby down sampling the data to a final storage rate of $f = f_{\mathrm{raw}} / M$. The following parameters were used:
\begin{itemize}
    \item For PA4: $N = 64$, $M = 26$.
    \item For PA5 and PA6: $N = 55$, $M = 23$.
\end{itemize}
With these parameters, one derives the filter 3\,dB cutoffs and sampling rates listed in Table~\ref{tab:array_properties}.

\section{Impact of pulse profiles}
\label{app:impacts-pulse-profiles}
Our main analysis employed a simple boxcar pulse profile (Equation~\ref{eq:pulsar-profile}) since the morphology of the Crab Pulsar's emission pulse at millimeter wavelengths is unknown. To validate that this choice does not significantly bias our period-averaged flux estimate, we tested our methodology using an alternative pulse shape: the von Mises profile \citep{Weltevrede:2008:profile}. This smooth, bell-shaped profile is commonly used in pulsar astronomy and is defined as
\begin{equation}
\label{eq:von mises}
f(\phi | \mu, c) = \exp(c \cos(\phi - \mu))/(2\pi I_0(c)),
\end{equation}
where $\phi$ is the pulse phase, $\mu$ is the pulse phase offset, $I(x)$ is the modified Bessel function of the first kind, and $c$ is the concentration parameter, which is related to the full width at half maximum of the pulse relative to its period, $D$, via the relation $c = (\log 2)/(2\sin^{2}(\pi D / 2))$. For this test, We set $D=1/10$ to approximately match the width of our boxcar bins.

We injected a simulated pulsar signal with a von Mises profile and a peak flux density of 100\,mJy. This leads to a period-averaged flux density of around 10.7\,mJy. This simulated source was injected at an off-target location (0.5$^\circ$ to the east) to avoid contamination from the Crab Nebula while preserving the noise statistics of the Crab observations. We then fitted the simulated data using both the standard boxcar templates and with the matched von Mises templates using the TOD-based method, which can easily accommodate alternative search templates.

\begin{figure}[t]
    \centering
    \includegraphics[width=0.6\linewidth]{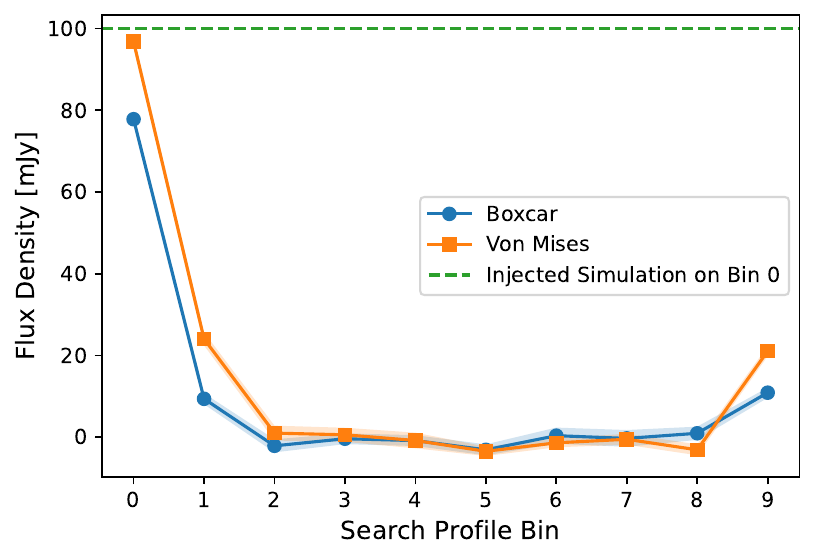}
    \caption{Comparison of flux recovery using boxcar versus von Mises search templates applied to an injected 100-mJy von Mises pulse signal at an off-target location. The boxcar template (blue) does not fully recover the injected flux in any individual phase bin due to profile mismatch, but yields an unbiased period-averaged flux of $9.2 \pm 0.6$ mJy. The matched von Mises template (orange) recovers $96.8 \pm 1.4$ mJy in the first phase bin but shows significant spillover to adjacent bins due to non-orthogonality, resulting in a biased period-averaged flux measurement of $13.4 \pm 0.6$ mJy.}
    \label{fig:flux_recovery_vonMises}
\end{figure}

Figure~\ref{fig:flux_recovery_vonMises} shows the results of the two fitting approaches. As expected, the boxcar template fails to fully recover the injected flux in any individual phase bin due to the mismatch with the true pulse profile. When computing the period-averaged flux density across all phase bins, we obtain $9.2 \pm 0.6$ mJy, which agrees with the injected period-averaged flux of 10.7\,mJy within 2.5$\sigma$ uncertainty. This acceptable agreement confirms that our period-averaged flux estimate is robust and relatively insensitive to the specific pulse profile morphology, validating our approach for the Crab Pulsar where the millimeter-wave pulse morphology remains unknown.

We then fitted the same simulation using a set of von Mises templates matching the input signal shape, centered at each of the 10 phase bins. As a result, we recover $96.8 \pm 1.4$ mJy in the zeroth phase bin, showing better sensitivity and less bias to the peak flux measurement when the template is properly matched. However, unlike the orthogonal boxcar profiles used in our main analysis, the von Mises profiles centered at different phase bins have non-zero overlaps. This causes the measurements to spill over into adjacent bins, as can be seen in Figure~\ref{fig:flux_recovery_vonMises}. The spillover effect leads to a biased period-averaged flux measurement of $13.4 \pm 0.6$ mJy, representing a 4.5$\sigma$ discrepancy from the injected period-averaged flux of 10.7\,mJy. This highlights the importance of using an orthogonal set of profile templates in future pulsar searches.

\bibliography{cite}{}

\begin{thebibliography}{}
\expandafter\ifx\csname natexlab\endcsname\relax\def\natexlab#1{#1}\fi
\providecommand{\url}[1]{\href{#1}{#1}}
\providecommand{\dodoi}[1]{doi:~\href{http://doi.org/#1}{\nolinkurl{#1}}}
\providecommand{\doeprint}[1]{\href{http://ascl.net/#1}{\nolinkurl{http://ascl.net/#1}}}
\providecommand{\doarXiv}[1]{\href{https://arxiv.org/abs/#1}{\nolinkurl{https://arxiv.org/abs/#1}}}

\bibitem[{{Abitbol} {et~al.}(2025){Abitbol}, {Abril-Cabezas}, {Adachi}, {Ade},
  {Adler}, {Agrawal}, {Aguirre}, {Ahmed}, {Aiola}, {Alford}, {Ali}, {Alonso},
  {Alvarez}, {An}, {Arnold}, {Ashton}, {Atkins}, {Austermann}, {Azzoni},
  {Baccigalupi}, {Baleato Lizancos}, {Barron}, {Barry}, {Bartlett},
  {Battaglia}, {Battye}, {Baxter}, {Bazarko}, {Beall}, {Bean}, {Beck},
  {Beckman}, {Begin}, {Beheshti}, {Beringue}, {Bhandarkar}, {Bhimani},
  {Bianchini}, {Biermann}, {Biquard}, {Bixler}, {Boada}, {Boettger}, {Bolliet},
  {Bond}, {Borrill}, {Borrow}, {Braithwaite}, {Brien}, {Brown}, {Bruno},
  {Bryan}, {Bustos}, {Cai}, {Calabrese}, {Calafut}, {Carl}, {Carones},
  {Carron}, {Challinor}, {Chanial}, {Chen}, {Cheung}, {Chiang}, {Chinone},
  {Chluba}, {Cho}, {Choi}, {Chu}, {Clancy}, {Clark}, {Clarke}, {Cleary},
  {Clements}, {Connors}, {Contaldi}, {Coppi}, {Corbett}, {Cothard}, {Coulton},
  {Crowley}, {Crowley}, {Cukierman}, {D'Ewart}, {Dachlythra}, {Datta},
  {Day-Weiss}, {de Haan}, {Devlin}, {Di Mascolo}, {Dicker}, {Dober}, {Doux},
  {Dow}, {Doyle}, {Duell}, {Duff}, {Duivenvoorden}, {Dunkley}, {Dutcher},
  {D{\"u}nner}, {Edenton}, {El Bouhargani}, {Errard}, {Fabbian}, {Fanfani},
  {Farren}, {Fergusson}, {Ferraro}, {Flauger}, {Foster}, {Freese}, {Frisch},
  {Frolov}, {Fuller}, {Galitzki}, {Gallardo}, {Galvez Ghersi}, {Ganga}, {Gao},
  {Garrido}, {Gawiser}, {Gerbino}, {Gerras}, {Giardiello}, {Gill}, {Gilles},
  {Giri}, {Gleave}, {Gluscevic}, {Goeckner-Wald}, {Golec}, {Gordon}, {Gralla},
  {Gratton}, {Green}, {Groh}, {Groppi}, {Guan}, {Gupta}, {Gudmundsson},
  {Hagstotz}, {Hargrave}, {Haridas}, {Harrington}, {Harrison}, {Hasegawa},
  {Hasselfield}, {Haynes}, {Hazumi}, {He}, {Healy}, {Henderson}, {Hensley},
  {Hertig}, {Herv{\'\i}as-Caimapo}, {Higuchi}, {Hill}, {Hill}, {Hilton},
  {Hilton}, {Hincks}, {Hinshaw}, {Hlo{\v{z}}ek}, {Ho}, {Ho}, {Ho}, {Hoang},
  {Hoh}, {Hornecker}, {Hornsby}, {Hotinli}, {Huang}, {Huber}, {Hubmayr},
  {Huffenberger}, {Hughes}, {Idicherian Lonappan}, {Ikape}, {Irwin}, {Iuliano},
  {Jaffe}, {Jain}, {Jense}, {Jeong}, {Johnson}, {Johnson}, {Johnson}, {Jones},
  {Jost}, {Kaneko}, {Karpel}, {Kasai}, {Katayama}, {Keating}, {Keller},
  {Keskitalo}, {Kim}, {Kisner}, \& {Kiuchi}}]{SO:SO_Collaboration:2025}
{Abitbol}, M., {Abril-Cabezas}, I., {Adachi}, S., {et~al.} 2025, \jcap, 2025,
  034, \dodoi{10.1088/1475-7516/2025/08/034}

\bibitem[{{Ade} {et~al.}(2019){Ade}, {Aguirre}, {Ahmed}, {Aiola}, {Ali},
  {Alonso}, {Alvarez}, {Arnold}, {Ashton}, {Austermann}, {Awan}, {Baccigalupi},
  {Baildon}, {Barron}, {Battaglia}, {Battye}, {Baxter}, {Bazarko}, {Beall},
  {Bean}, {Beck}, {Beckman}, {Beringue}, {Bianchini}, {Boada}, {Boettger},
  {Bond}, {Borrill}, {Brown}, {Bruno}, {Bryan}, {Calabrese}, {Calafut},
  {Calisse}, {Carron}, {Challinor}, {Chesmore}, {Chinone}, {Chluba}, {Cho},
  {Choi}, {Coppi}, {Cothard}, {Coughlin}, {Crichton}, {Crowley}, {Crowley},
  {Cukierman}, {D'Ewart}, {D{\"u}nner}, {de Haan}, {Devlin}, {Dicker},
  {Didier}, {Dobbs}, {Dober}, {Duell}, {Duff}, {Duivenvoorden}, {Dunkley},
  {Dusatko}, {Errard}, {Fabbian}, {Feeney}, {Ferraro}, {Flux{\`a}}, {Freese},
  {Frisch}, {Frolov}, {Fuller}, {Fuzia}, {Galitzki}, {Gallardo}, {Tomas Galvez
  Ghersi}, {Gao}, {Gawiser}, {Gerbino}, {Gluscevic}, {Goeckner-Wald}, {Golec},
  {Gordon}, {Gralla}, {Green}, {Grigorian}, {Groh}, {Groppi}, {Guan},
  {Gudmundsson}, {Han}, {Hargrave}, {Hasegawa}, {Hasselfield}, {Hattori},
  {Haynes}, {Hazumi}, {He}, {Healy}, {Henderson}, {Hervias-Caimapo}, {Hill},
  {Hill}, {Hilton}, {Hilton}, {Hincks}, {Hinshaw}, {Hlo{\v{z}}ek}, {Ho}, {Ho},
  {Howe}, {Huang}, {Hubmayr}, {Huffenberger}, {Hughes}, {Ijjas}, {Ikape},
  {Irwin}, {Jaffe}, {Jain}, {Jeong}, {Kaneko}, {Karpel}, {Katayama}, {Keating},
  {Kernasovskiy}, {Keskitalo}, {Kisner}, {Kiuchi}, {Klein}, {Knowles},
  {Koopman}, {Kosowsky}, {Krachmalnicoff}, {Kuenstner}, {Kuo}, {Kusaka},
  {Lashner}, {Lee}, {Lee}, {Leon}, {Leung}, {Lewis}, {Li}, {Li}, {Limon},
  {Linder}, {Lopez-Caraballo}, {Louis}, {Lowry}, {Lungu}, {Madhavacheril},
  {Mak}, {Maldonado}, {Mani}, {Mates}, {Matsuda}, {Maurin}, {Mauskopf}, {May},
  {McCallum}, {McKenney}, {McMahon}, {Meerburg}, {Meyers}, {Miller},
  {Mirmelstein}, {Moodley}, {Munchmeyer}, {Munson}, {Naess}, {Nati},
  {Navaroli}, {Newburgh}, {Nguyen}, {Niemack}, {Nishino}, {Orlowski-Scherer},
  {Page}, {Partridge}, {Peloton}, {Perrotta}, {Piccirillo}, {Pisano},
  {Poletti}, {Puddu}, {Puglisi}, {Raum}, {Reichardt}, {Remazeilles},
  {Rephaeli}, {Riechers}, {Rojas}, {Roy}, {Sadeh}, {Sakurai}, {Salatino},
  {Sathyanarayana Rao}, {Schaan}, {Schmittfull}, {Sehgal}, \&
  {Seibert}}]{SO:SO_Collaboration:2019}
{Ade}, P., {Aguirre}, J., {Ahmed}, Z., {et~al.} 2019, \jcap, 2019, 056,
  \dodoi{10.1088/1475-7516/2019/02/056}

\bibitem[{{Aumont} {et~al.}(2010){Aumont}, {Conversi}, {Thum}, {Wiesemeyer},
  {Falgarone}, {Mac{\'\i}as-P{\'e}rez}, {Piacentini}, {Pointecouteau},
  {Ponthieu}, {Puget}, {Rosset}, {Tauber}, \& {Tristram}}]{Aumont:2010}
{Aumont}, J., {Conversi}, L., {Thum}, C., {et~al.} 2010, \aap, 514, A70,
  \dodoi{10.1051/0004-6361/200913834}

\bibitem[{{Battistelli} {et~al.}(2008{\natexlab{a}}){Battistelli}, {Amiri},
  {Burger}, {Devlin}, {Dicker}, {Doriese}, {D{\"u}nner}, {Fisher}, {Fowler},
  {Halpern}, {Hasselfield}, {Hilton}, {Hincks}, {Irwin}, {Kaul}, {Klein},
  {Knotek}, {Lau}, {Limon}, {Marriage}, {Niemack}, {Page}, {Reintsema},
  {Staggs}, {Swetz}, {Switzer}, {Thornton}, \& {Zhao}}]{ACT:Battistelli:2008}
{Battistelli}, E.~S., {Amiri}, M., {Burger}, B., {et~al.} 2008{\natexlab{a}},
  in Society of Photo-Optical Instrumentation Engineers (SPIE) Conference
  Series, Vol. 7020, Millimeter and Submillimeter Detectors and Instrumentation
  for Astronomy IV, ed. W.~D. {Duncan}, W.~S. {Holland}, S.~{Withington}, \&
  J.~{Zmuidzinas}, 702028, \dodoi{10.1117/12.789738}

\bibitem[{{Battistelli} {et~al.}(2008{\natexlab{b}}){Battistelli}, {Amiri},
  {Burger}, {Halpern}, {Knotek}, {Ellis}, {Gao}, {Kelly}, {Macintosh}, {Irwin},
  \& {Reintsema}}]{ACT:Battistelli:2008b}
{Battistelli}, E.~S., {Amiri}, M., {Burger}, B., {et~al.} 2008{\natexlab{b}},
  Journal of Low Temperature Physics, 151, 908,
  \dodoi{10.1007/s10909-008-9772-z}

\bibitem[{{Biermann} {et~al.}(2025){Biermann}, {Li}, {Naess}, {Choi}, {Clark},
  {Devlin}, {Dunkley}, {Gallardo}, {Guan}, {Foster}, {Hasselfield},
  {Herv{\'\i}as-Caimapo}, {Hilton}, {Hincks}, {Ho}, {Hood}, {Huffenberger},
  {Kosowsky}, {Niemack}, {Orlowski-Scherer}, {Page}, {Partridge}, {Salatino},
  {Sif{\'o}n}, {Staggs}, {Vargas}, \& {Wollack}}]{ACT:Biermann:2025}
{Biermann}, E., {Li}, Y., {Naess}, S., {et~al.} 2025, \apj, 986, 7,
  \dodoi{10.3847/1538-4357/adce70}

\bibitem[{{Blandford}(2002)}]{Blandford:2002}
{Blandford}, R.~D. 2002, in Lighthouses of the Universe: The Most Luminous
  Celestial Objects and Their Use for Cosmology, ed. M.~{Gilfanov},
  R.~{Sunyeav}, \& E.~{Churazov}, 381, \dodoi{10.1007/10856495_59}

\bibitem[{{B{\"u}hler} \& {Blandford}(2014)}]{Buhler:2014}
{B{\"u}hler}, R., \& {Blandford}, R. 2014, Reports on Progress in Physics, 77,
  066901, \dodoi{10.1088/0034-4885/77/6/066901}

\bibitem[{{Camilo} {et~al.}(2008){Camilo}, {Reynolds}, {Johnston}, {Halpern},
  \& {Ransom}}]{Camilo:2008}
{Camilo}, F., {Reynolds}, J., {Johnston}, S., {Halpern}, J.~P., \& {Ransom},
  S.~M. 2008, \apj, 679, 681, \dodoi{10.1086/587054}

\bibitem[{{Camilo} {et~al.}(2007){Camilo}, {Ransom}, {Pe{\~n}alver},
  {Karastergiou}, {van Kerkwijk}, {Durant}, {Halpern}, {Reynolds}, {Thum},
  {Helfand}, {Zimmerman}, \& {Cognard}}]{Camilo:2007}
{Camilo}, F., {Ransom}, S.~M., {Pe{\~n}alver}, J., {et~al.} 2007, \apj, 669,
  561, \dodoi{10.1086/521548}

\bibitem[{{CCAT-Prime Collaboration} {et~al.}(2023){CCAT-Prime Collaboration},
  {Aravena}, {Austermann}, {Basu}, {Battaglia}, {Beringue}, {Bertoldi},
  {Bigiel}, {Bond}, {Breysse}, {Broughton}, {Bustos}, {Chapman}, {Charmetant},
  {Choi}, {Chung}, {Clark}, {Cothard}, {Crites}, {Dev}, {Douglas}, {Duell},
  {D{\"u}nner}, {Ebina}, {Erler}, {Fich}, {Fissel}, {Foreman}, {Freundt},
  {Gallardo}, {Gao}, {Garc{\'\i}a}, {Giovanelli}, {Golec}, {Groppi}, {Haynes},
  {Henke}, {Hensley}, {Herter}, {Higgins}, {Hlo{\v{z}}ek}, {Huber}, {Huber},
  {Hubmayr}, {Jackson}, {Johnstone}, {Karoumpis}, {Keating}, {Komatsu}, {Li},
  {Magnelli}, {Matthews}, {Mauskopf}, {McMahon}, {Meerburg}, {Meyers},
  {Muralidhara}, {Murray}, {Niemack}, {Nikola}, {Okada}, {Puddu}, {Riechers},
  {Rosolowsky}, {Rossi}, {Rotermund}, {Roy}, {Sadavoy}, {Schaaf}, {Schilke},
  {Scott}, {Simon}, {Sinclair}, {Sivakoff}, {Stacey}, {Stutz}, {Stutzki},
  {Tahani}, {Thanjavur}, {Timmermann}, {Ullom}, {van Engelen}, {Vavagiakis},
  {Vissers}, {Wheeler}, {White}, {Zhu}, \&
  {Zou}}]{CCAT-PrimeCollaboration:2023}
{CCAT-Prime Collaboration}, {Aravena}, M., {Austermann}, J.~E., {et~al.} 2023,
  \apjs, 264, 7, \dodoi{10.3847/1538-4365/ac9838}

\bibitem[{{Chichura} {et~al.}(2022){Chichura}, {Foster}, {Patel}, {Ossa-Jaen},
  {Ade}, {Ahmed}, {Anderson}, {Archipley}, {Austermann}, {Avva}, {Balkenhol},
  {Barry}, {Thakur}, {Beall}, {Benabed}, {Bender}, {Benson}, {Bianchini},
  {Bleem}, {Bouchet}, {Bryant}, {Byrum}, {Carlstrom}, {Carter}, {Cecil},
  {Chang}, {Chaubal}, {Chen}, {Chiang}, {Cho}, {Chou}, {Citron}, {Cliche},
  {Crawford}, {Crites}, {Cukierman}, {Daley}, {Denison}, {Dibert}, {Ding},
  {Dobbs}, {Dutcher}, {Everett}, {Feng}, {Ferguson}, {Fu}, {Galli},
  {Gallicchio}, {Gambrel}, {Gardner}, {George}, {Goeckner-Wald}, {Gualtieri},
  {Guns}, {Gupta}, {Guyser}, {de Haan}, {Halverson}, {Harke-Hosemann},
  {Harrington}, {Henning}, {Hilton}, {Hivon}, {Holder}, {Holzapfel}, {Hood},
  {Howe}, {Hrubes}, {Huang}, {Hubmayr}, {Irwin}, {Jeong}, {Jonas}, {Jones},
  {Khaire}, {Knox}, {Kofman}, {Korman}, {Kubik}, {Kuhlmann}, {Kuo}, {Lee},
  {Leitch}, {Li}, {Lowitz}, {Lu}, {Marrone}, {McMahon}, {Meyer}, {Michalik},
  {Millea}, {Mocanu}, {Montgomery}, {Moran}, {Nadolski}, {Natoli}, {Nguyen},
  {Nibarger}, {Noble}, {Novosad}, {Omori}, {Padin}, {Pan}, {Paschos}, {Patil},
  {Pearson}, {Phadke}, {Posada}, {Prabhu}, {Pryke}, {Quan}, {Rahlin},
  {Reichardt}, {Riebel}, {Riedel}, {Rouble}, {Ruhl}, {Saliwanchik}, {Sayre},
  {Schaffer}, {Schiappucci}, {Shirokoff}, {Sievers}, {Smecher}, {Sobrin},
  {Springmann}, {Stark}, {Stephen}, {Story}, {Suzuki}, {Tandoi}, {Thompson},
  {Thorne}, {Tucker}, {Umilta}, {Vale}, {Veach}, {Vieira}, {Wang}, {Whitehorn},
  {Wu}, {Yefremenko}, {Yoon}, \& {Young}}]{SPT:Chichura:2022}
{Chichura}, P.~M., {Foster}, A., {Patel}, C., {et~al.} 2022, \apj, 936, 173,
  \dodoi{10.3847/1538-4357/ac89ec}

\bibitem[{{Chown} {et~al.}(2018){Chown}, {Omori}, {Aylor}, {Benson}, {Bleem},
  {Carlstrom}, {Chang}, {Cho}, {Crawford}, {Crites}, {de Haan}, {Dobbs},
  {Everett}, {George}, {Henning}, {Halverson}, {Harrington}, {Holder},
  {Holzapfel}, {Hou}, {Hrubes}, {Knox}, {Lee}, {Luong-Van}, {Marrone},
  {McMahon}, {Meyer}, {Millea}, {Mocanu}, {Mohr}, {Natoli}, {Padin}, {Pryke},
  {Reichardt}, {Ruhl}, {Sayre}, {Schaffer}, {Shirokoff}, {Simard},
  {Staniszewski}, {Stark}, {Story}, {Vanderlinde}, {Vieira}, {Williamson},
  {Wu}, \& {South Pole Telescope Collaboration}}]{Chown:2018}
{Chown}, R., {Omori}, Y., {Aylor}, K., {et~al.} 2018, \apjs, 239, 10,
  \dodoi{10.3847/1538-4365/aae694}

\bibitem[{{Chu} {et~al.}(2021){Chu}, {Ng}, {Kong}, \& {Chang}}]{Chu:2021}
{Chu}, C.-Y., {Ng}, C.~Y., {Kong}, A. K.~H., \& {Chang}, H.-K. 2021, \mnras,
  503, 1214, \dodoi{10.1093/mnras/stab349}

\bibitem[{{Cordes} \& {Lazio}(1997)}]{Cordes:1997}
{Cordes}, J.~M., \& {Lazio}, T. J.~W. 1997, \apj, 475, 557,
  \dodoi{10.1086/303569}

\bibitem[{{Foster} {et~al.}(2025){Foster}, {Chokshi}, {Anderson},
  {Ansarinejad}, {Archipley}, {Balkenhol}, {Benabed}, {Bender}, {Barron},
  {Benson}, {Bianchini}, {Bleem}, {Bouchet}, {Bryant}, {Camphuis}, {Carlstrom},
  {Chang}, {Chaubal}, {Chichura}, {Chou}, {Coerver}, {Crawford}, {Daley}, {de
  Haan}, {Dibert}, {Dobbs}, {Doussot}, {Dutcher}, {Everett}, {Feng},
  {Ferguson}, {Fichman}, {Galli}, {Gambrel}, {Gardner}, {Ge}, {Goeckner-Wald},
  {Gualtieri}, {Guidi}, {Guns}, {Halverson}, {Hivon}, {Holder}, {Holzapfel},
  {Hood}, {Hryciuk}, {Huang}, {K{\'e}ruzor{\'e}}, {Khalife}, {Knox},
  {Kornoelje}, {Korman}, {Kuo}, {Levy}, {Lowitz}, {Lu}, {Maniyar}, {Martsen},
  {Menanteau}, {Millea}, {Montgomery}, {Nakato}, {Natoli}, {Noble}, {Omori},
  {Pan}, {Paschos}, {Phadke}, {Pollak}, {Prabhu}, {Quan}, {Rahimi}, {Rahlin},
  {Reichardt}, {Rouble}, {Ruhl}, {Schiappucci}, {Sobrin}, {Stark}, {Stephen},
  {Tandoi}, {Thorne}, {Trendafilova}, {Umilta}, {Vieira}, {Vitrier}, {Wan},
  {Whitehorn}, {Wu}, {Young}, \& {Zebrowski}}]{SPT:Foster:2025}
{Foster}, A., {Chokshi}, A., {Anderson}, A.~J., {et~al.} 2025, The Open Journal
  of Astrophysics, 8, 51, \dodoi{10.33232/001c.137526}

\bibitem[{{Gold}(1968)}]{Gold:1968}
{Gold}, T. 1968, \nat, 218, 731, \dodoi{10.1038/218731a0}

\bibitem[{{Guns} {et~al.}(2021){Guns}, {Foster}, {Daley}, {Rahlin},
  {Whitehorn}, {Ade}, {Ahmed}, {Anderes}, {Anderson}, {Archipley}, {Avva},
  {Aylor}, {Balkenhol}, {Barry}, {Basu Thakur}, {Benabed}, {Bender}, {Benson},
  {Bianchini}, {Bleem}, {Bouchet}, {Bryant}, {Byrum}, {Carlstrom}, {Carter},
  {Cecil}, {Chang}, {Chaubal}, {Chen}, {Cho}, {Chou}, {Cliche}, {Crawford},
  {Cukierman}, {de Haan}, {Denison}, {Dibert}, {Ding}, {Dobbs}, {Dutcher},
  {Everett}, {Feng}, {Ferguson}, {Fu}, {Galli}, {Gambrel}, {Gardner},
  {Goeckner-Wald}, {Gualtieri}, {Gupta}, {Guyser}, {Halverson},
  {Harke-Hosemann}, {Harrington}, {Henning}, {Hilton}, {Hivon}, {Holder},
  {Holzapfel}, {Hood}, {Howe}, {Huang}, {Irwin}, {Jeong}, {Jonas}, {Jones},
  {Khaire}, {Knox}, {Kofman}, {Korman}, {Kubik}, {Kuhlmann}, {Kuo}, {Lee},
  {Leitch}, {Lowitz}, {Lu}, {Marrone}, {Meyer}, {Michalik}, {Millea},
  {Montgomery}, {Nadolski}, {Natoli}, {Nguyen}, {Noble}, {Novosad}, {Omori},
  {Padin}, {Pan}, {Paschos}, {Pearson}, {Phadke}, {Posada}, {Prabhu}, {Quan},
  {Reichardt}, {Riebel}, {Riedel}, {Rouble}, {Ruhl}, {Sayre}, {Schiappucci},
  {Shirokoff}, {Smecher}, {Sobrin}, {Stark}, {Stephen}, {Story}, {Suzuki},
  {Thompson}, {Thorne}, {Tucker}, {Umilta}, {Vale}, {Vieira}, {Wang}, {Wu},
  {Yefremenko}, {Yoon}, {Young}, \& {Zhang}}]{SPT:Guns:2021}
{Guns}, S., {Foster}, A., {Daley}, C., {et~al.} 2021, \apj, 916, 98,
  \dodoi{10.3847/1538-4357/ac06a3}

\bibitem[{{Hankins} {et~al.}(2015){Hankins}, {Jones}, \&
  {Eilek}}]{Hankins:2015}
{Hankins}, T.~H., {Jones}, G., \& {Eilek}, J.~A. 2015, \apj, 802, 130,
  \dodoi{10.1088/0004-637X/802/2/130}

\bibitem[{{Hasselfield}(2013)}]{Thesis:Hasselfield:2013}
{Hasselfield}, M. 2013, PhD thesis, University of British Columbia

\bibitem[{{Henderson} {et~al.}(2016){Henderson}, {Allison}, {Austermann},
  {Baildon}, {Battaglia}, {Beall}, {Becker}, {De Bernardis}, {Bond},
  {Calabrese}, {Choi}, {Coughlin}, {Crowley}, {Datta}, {Devlin}, {Duff},
  {Dunkley}, {D{\"u}nner}, {van Engelen}, {Gallardo}, {Grace}, {Hasselfield},
  {Hills}, {Hilton}, {Hincks}, {Hloẑek}, {Ho}, {Hubmayr}, {Huffenberger},
  {Hughes}, {Irwin}, {Koopman}, {Kosowsky}, {Li}, {McMahon}, {Munson}, {Nati},
  {Newburgh}, {Niemack}, {Niraula}, {Page}, {Pappas}, {Salatino}, {Schillaci},
  {Schmitt}, {Sehgal}, {Sherwin}, {Sievers}, {Simon}, {Spergel}, {Staggs},
  {Stevens}, {Thornton}, {Van Lanen}, {Vavagiakis}, {Ward}, \&
  {Wollack}}]{ACT:Henderson:2016}
{Henderson}, S.~W., {Allison}, R., {Austermann}, J., {et~al.} 2016, Journal of
  Low Temperature Physics, 184, 772, \dodoi{10.1007/s10909-016-1575-z}

\bibitem[{{Herv{\'\i}as-Caimapo} {et~al.}(2024){Herv{\'\i}as-Caimapo}, {Naess},
  {Hincks}, {Calabrese}, {Devlin}, {Dunkley}, {D{\"u}nner}, {Gallardo},
  {Hilton}, {Ho}, {Huffenberger}, {Ma}, {Madhavacheril}, {Niemack},
  {Orlowski-Scherer}, {Page}, {Partridge}, {Puddu}, {Salatino}, {Sif{\'o}n},
  {Staggs}, {Vargas}, {Vavagiakis}, \& {Wollack}}]{ACT:Hervias-Caimapo:2024}
{Herv{\'\i}as-Caimapo}, C., {Naess}, S., {Hincks}, A.~D., {et~al.} 2024,
  \mnras, 529, 3020, \dodoi{10.1093/mnras/stae583}

\bibitem[{{Koopman} {et~al.}(2018){Koopman}, {Cothard}, {Choi}, {Crowley},
  {Duff}, {Henderson}, {Ho}, {Hubmayr}, {Gallardo}, {Nati}, {Niemack}, {Simon},
  {Staggs}, {Stevens}, {Vavagiakis}, \& {Wollack}}]{ACT:Koopman:2018}
{Koopman}, B.~J., {Cothard}, N.~F., {Choi}, S.~K., {et~al.} 2018, Journal of
  Low Temperature Physics, 193, 1103, \dodoi{10.1007/s10909-018-1957-5}

\bibitem[{{Kramer} {et~al.}(1997{\natexlab{a}}){Kramer}, {Jessner},
  {Doroshenko}, \& {Wielebinski}}]{Kramer:1997:7mm}
{Kramer}, M., {Jessner}, A., {Doroshenko}, O., \& {Wielebinski}, R.
  1997{\natexlab{a}}, \apj, 488, 364, \dodoi{10.1086/304706}

\bibitem[{{Kramer} {et~al.}(1996){Kramer}, {Xilouris}, {Jessner},
  {Wielebinski}, \& {Timofeev}}]{Kramer:1996}
{Kramer}, M., {Xilouris}, K.~M., {Jessner}, A., {Wielebinski}, R., \&
  {Timofeev}, M. 1996, \aap, 306, 867

\bibitem[{{Kramer} {et~al.}(1997{\natexlab{b}}){Kramer}, {Xilouris}, \&
  {Rickett}}]{Kramer:1997}
{Kramer}, M., {Xilouris}, K.~M., \& {Rickett}, B. 1997{\natexlab{b}}, \aap,
  321, 513

\bibitem[{{Kuiper} {et~al.}(2001){Kuiper}, {Hermsen}, {Cusumano}, {Diehl},
  {Sch{\"o}nfelder}, {Strong}, {Bennett}, \& {McConnell}}]{Kuiper:2001}
{Kuiper}, L., {Hermsen}, W., {Cusumano}, G., {et~al.} 2001, \aap, 378, 918,
  \dodoi{10.1051/0004-6361:20011256}

\bibitem[{{Li} {et~al.}(2023){Li}, {Biermann}, {Naess}, {Aiola}, {An},
  {Battaglia}, {Bhandarkar}, {Calabrese}, {Choi}, {Crowley}, {Devlin}, {Duell},
  {Duff}, {Dunkley}, {D{\"u}nner}, {Gallardo}, {Guan}, {Herv{\'\i}as-Caimapo},
  {Hincks}, {Hubmayr}, {Huffenberger}, {Hughes}, {Kosowsky}, {Louis},
  {Mallaby-Kay}, {McMahon}, {Nati}, {Niemack}, {Orlowski-Scherer}, {Page},
  {Salatino}, {Sif{\'o}n}, {Staggs}, {Vargas}, {Vavagiakis}, {Wang}, \&
  {Wollack}}]{ACT:Li:2023}
{Li}, Y., {Biermann}, E., {Naess}, S., {et~al.} 2023, \apj, 956, 36,
  \dodoi{10.3847/1538-4357/ace599}

\bibitem[{{Lyne} {et~al.}(1993){Lyne}, {Pritchard}, \& {Graham
  Smith}}]{Lyne:1993:crab}
{Lyne}, A.~G., {Pritchard}, R.~S., \& {Graham Smith}, F. 1993, \mnras, 265,
  1003, \dodoi{10.1093/mnras/265.4.1003}

\bibitem[{{Macquart} \& {Kanekar}(2015)}]{Macquart:2015}
{Macquart}, J.-P., \& {Kanekar}, N. 2015, \apj, 805, 172,
  \dodoi{10.1088/0004-637X/805/2/172}

\bibitem[{{Melrose}(2017)}]{Melrose:2017}
{Melrose}, D.~B. 2017, Reviews of Modern Plasma Physics, 1, 5,
  \dodoi{10.1007/s41614-017-0007-0}

\bibitem[{{Meyer} {et~al.}(2010){Meyer}, {Horns}, \& {Zechlin}}]{Meyer:2010}
{Meyer}, M., {Horns}, D., \& {Zechlin}, H.~S. 2010, \aap, 523, A2,
  \dodoi{10.1051/0004-6361/201014108}

\bibitem[{{Michel}(1978)}]{Michel:1978}
{Michel}, F.~C. 1978, \apj, 220, 1101, \dodoi{10.1086/155995}

\bibitem[{{Mignani} {et~al.}(2017){Mignani}, {Paladino}, {Rudak}, {Zajczyk},
  {Corongiu}, {de Luca}, {Hummel}, {Possenti}, {Geppert}, {Burgay}, \&
  {Marconi}}]{Mignani:2017}
{Mignani}, R.~P., {Paladino}, R., {Rudak}, B., {et~al.} 2017, \apjl, 851, L10,
  \dodoi{10.3847/2041-8213/aa9c3e}

\bibitem[{{Morris} {et~al.}(1997){Morris}, {Kramer}, {Thum}, {Wielebinski},
  {Grewing}, {Penalver}, {Jessner}, {Butin}, \& {Brunswig}}]{Morris:1997}
{Morris}, D., {Kramer}, M., {Thum}, C., {et~al.} 1997, \aap, 322, L17

\bibitem[{{Naess} {et~al.}(2021{\natexlab{a}}){Naess}, {Battaglia}, {Richard
  Bond}, {Calabrese}, {Choi}, {Cothard}, {Devlin}, {Duell}, {Duivenvoorden},
  {Dunkley}, {D{\"u}nner}, {Gallardo}, {Gralla}, {Guan}, {Halpern}, {Colin
  Hill}, {Hilton}, {Huffenberger}, {Koopman}, {Kosowsky}, {Madhavacheril},
  {McMahon}, {Nati}, {Niemack}, {Page}, {Partridge}, {Salatino}, {Sehgal},
  {Spergel}, {Staggs}, {Wollack}, \& {Xu}}]{ACT:Naess:2021}
{Naess}, S., {Battaglia}, N., {Richard Bond}, J., {et~al.} 2021{\natexlab{a}},
  \apj, 915, 14, \dodoi{10.3847/1538-4357/abfe6d}

\bibitem[{{Naess} {et~al.}(2021{\natexlab{b}}){Naess}, {Aiola}, {Battaglia},
  {Bond}, {Calabrese}, {Choi}, {Cothard}, {Halpern}, {Hill}, {Koopman},
  {Devlin}, {McMahon}, {Dicker}, {Duivenvoorden}, {Dunkley}, {Fanfani},
  {Ferraro}, {Gallardo}, {Guan}, {Han}, {Hasselfield}, {Hincks},
  {Huffenberger}, {Kosowsky}, {Louis}, {Macinnis}, {Madhavacheril}, {Nati},
  {Niemack}, {Page}, {Salatino}, {Schaan}, {Orlowski-Scherer}, {Schillaci},
  {Schmitt}, {Sehgal}, {Sif{\'o}n}, {Staggs}, {Engelen}, \&
  {Wollack}}]{ACT:Naess:2021:P9}
{Naess}, S., {Aiola}, S., {Battaglia}, N., {et~al.} 2021{\natexlab{b}}, \apj,
  923, 224, \dodoi{10.3847/1538-4357/ac2307}

\bibitem[{{Naess} {et~al.}(2025){Naess}, {Guan}, {Duivenvoorden},
  {Hasselfield}, {Wang}, {Abril-Cabezas}, {Addison}, {Ade}, {Aiola}, {Alford},
  {Alonso}, {Amiri}, {An}, {Atkins}, {Austermann}, {Barbavara}, {Battaglia},
  {Battistelli}, {Beall}, {Bean}, {Beheshti}, {Beringue}, {Bhandarkar},
  {Biermann}, {Bolliet}, {Bond}, {Calabrese}, {Capalbo}, {Carrero}, {Chen},
  {Chesmore}, {Cho}, {Choi}, {Clark}, {Cordova Rosado}, {Cothard}, {Coughlin},
  {Coulton}, {Crichton}, {Crowley}, {Devlin}, {Dicker}, {Duell}, {Duff},
  {Dunkley}, {Dunner}, {Embil Villagra}, {Fankhanel}, {Farren}, {Ferraro},
  {Foster}, {Freundt}, {Fuzia}, {Gallardo}, {Garrido}, {Giardiello}, {Gill},
  {Givans}, {Gluscevic}, {Golec}, {Gong}, {Halpern}, {Harrison}, {Healy},
  {Henderson}, {Hensley}, {Herv{\'\i}as-Caimapo}, {Hill}, {Hilton}, {Hilton},
  {Hincks}, {Hlo{\v{z}}ek}, {Ho}, {Hood}, {Hornecker}, {Huber}, {Hubmayr},
  {Huffenberger}, {Hughes}, {Ikape}, {Irwin}, {Isopi}, {Jense}, {Joshi},
  {Keller}, {Kim}, {Knowles}, {Koopman}, {Kosowsky}, {Kramer}, {Kusiak}, {La
  Posta}, {Lagu{\"e}}, {Lakey}, {Lee}, {Li}, {Li}, {Limon}, {Lokken}, {Louis},
  {Lungu}, {MacCrann}, {MacInnis}, {Madhavacheril}, {Maldonado}, {Maldonado},
  {Mallaby-Kay}, {Marques}, {van Marrewijk}, {McCarthy}, {McMahon}, {Mehta},
  {Menanteau}, {Moodley}, {Morris}, {Mroczkowski}, {Namikawa}, {Nati},
  {Nerval}, {Newburgh}, {Nicola}, {Niemack}, {Nolta}, {Orlowski-Scherer},
  {Page}, {Pandey}, {Partridge}, {Perez Sarmiento}, {Prince}, {Puddu}, {Qu},
  {Quiroga}, {Ragavan}, {Ried Guachalla}, {Rogers}, {Rojas}, {Sakuma},
  {Schaan}, {Schmitt}, {Sehgal}, {Shaikh}, {Sherwin}, {Sierra}, {Sievers},
  {Sif{\'o}n}, {Simon}, {Sonka}, {Spergel}, {Staggs}, {Storer}, {Surrao},
  {Switzer}, {Tampier}, {Thornton}, {Trac}, {Tucker}, {Ullom}, {Vale}, {Van
  Engelen}, {Van Lanen}, {Vargas}, {Vavagiakis}, {Wagoner}, {Wenzl}, {Wollack},
  \& {Zheng}}]{ACT:Naess:2025}
{Naess}, S., {Guan}, Y., {Duivenvoorden}, A.~J., {et~al.} 2025, arXiv e-prints,
  arXiv:2503.14451, \dodoi{10.48550/arXiv.2503.14451}

\bibitem[{{Orlowski-Scherer} {et~al.}(2024){Orlowski-Scherer}, {Venterea},
  {Battaglia}, {Naess}, {Bhandarkar}, {Biermann}, {Calabrese}, {Devlin},
  {Dunkley}, {Herv{\'\i}as-Caimapo}, {Gallardo}, {Hilton}, {Hincks}, {Knowles},
  {Li}, {McMahon}, {Niemack}, {Page}, {Partridge}, {Salatino}, {Sievers},
  {Sif{\'o}n}, {Staggs}, {van Engelen}, {Vargas}, {Vavagiakis}, \&
  {Wollack}}]{ACT:Orlowski-Scherer:2024}
{Orlowski-Scherer}, J., {Venterea}, R.~C., {Battaglia}, N., {et~al.} 2024,
  \apj, 964, 138, \dodoi{10.3847/1538-4357/ad21fe}

\bibitem[{{Page} {et~al.}(2007){Page}, {Hinshaw}, {Komatsu}, {Nolta},
  {Spergel}, {Bennett}, {Barnes}, {Bean}, {Dor{\'e}}, {Dunkley}, {Halpern},
  {Hill}, {Jarosik}, {Kogut}, {Limon}, {Meyer}, {Odegard}, {Peiris}, {Tucker},
  {Verde}, {Weiland}, {Wollack}, \& {Wright}}]{Page:2007}
{Page}, L., {Hinshaw}, G., {Komatsu}, E., {et~al.} 2007, \apjs, 170, 335,
  \dodoi{10.1086/513699}

\bibitem[{{Rocha} {et~al.}(2023){Rocha}, {Keskitalo}, {Partridge}, {Marscher},
  {O'Dea}, {Pearson}, \& {G{\'o}rski}}]{Rocha:2023}
{Rocha}, G., {Keskitalo}, R., {Partridge}, B., {et~al.} 2023, \aap, 669, A92,
  \dodoi{10.1051/0004-6361/202141995}

\bibitem[{{Sollerman} {et~al.}(2000){Sollerman}, {Lundqvist}, {Lindler},
  {Chevalier}, {Fransson}, {Gull}, {Pun}, \& {Sonneborn}}]{Sollerman:2000}
{Sollerman}, J., {Lundqvist}, P., {Lindler}, D., {et~al.} 2000, \apj, 537, 861,
  \dodoi{10.1086/309062}

\bibitem[{{Swetz} {et~al.}(2011){Swetz}, {Ade}, {Amiri}, {Appel},
  {Battistelli}, {Burger}, {Chervenak}, {Devlin}, {Dicker}, {Doriese},
  {D{\"u}nner}, {Essinger-Hileman}, {Fisher}, {Fowler}, {Halpern},
  {Hasselfield}, {Hilton}, {Hincks}, {Irwin}, {Jarosik}, {Kaul}, {Klein},
  {Lau}, {Limon}, {Marriage}, {Marsden}, {Martocci}, {Mauskopf}, {Moseley},
  {Netterfield}, {Niemack}, {Nolta}, {Page}, {Parker}, {Staggs}, {Stryzak},
  {Switzer}, {Thornton}, {Tucker}, {Wollack}, \& {Zhao}}]{ACT:Swetz:2011}
{Swetz}, D.~S., {Ade}, P.~A.~R., {Amiri}, M., {et~al.} 2011, \apjs, 194, 41,
  \dodoi{10.1088/0067-0049/194/2/41}

\bibitem[{{Tandoi} {et~al.}(2024){Tandoi}, {Guns}, {Foster}, {Ade}, {Anderson},
  {Ansarinejad}, {Archipley}, {Balkenhol}, {Benabed}, {Bender}, {Benson},
  {Bianchini}, {Bleem}, {Bouchet}, {Bryant}, {Camphuis}, {Carlstrom}, {Cecil},
  {Chang}, {Chaubal}, {Chichura}, {Chou}, {Coerver}, {Crawford}, {Cukierman},
  {Daley}, {de Haan}, {Dibert}, {Dobbs}, {Doussot}, {Dutcher}, {Everett},
  {Feng}, {Ferguson}, {Fichman}, {Galli}, {Gambrel}, {Gardner}, {Ge},
  {Goeckner-Wald}, {Gualtieri}, {Guidi}, {Halverson}, {Hivon}, {Holder},
  {Holzapfel}, {Hood}, {Huang}, {K{\'e}ruzor{\'e}}, {Knox}, {Korman},
  {Kornoelje}, {Kuo}, {Lee}, {Levy}, {Lowitz}, {Lu}, {Maniyar}, {Menanteau},
  {Millea}, {Montgomery}, {Moon}, {Nakato}, {Natoli}, {Noble}, {Novosad},
  {Omori}, {Padin}, {Pan}, {Paschos}, {Phadke}, {Prabhu}, {Qu}, {Quan},
  {Rahimi}, {Rahlin}, {Reichardt}, {Reuter}, {Rouble}, {Ruhl}, {Schiappucci},
  {Smecher}, {Sobrin}, {Stark}, {Stephen}, {Suzuki}, {Thompson}, {Thorne},
  {Trendafilova}, {Tucker}, {Umilta}, {Vieira}, {Wan}, {Wang}, {Whitehorn},
  {Wu}, {Yefremenko}, {Young}, \& {Zebrowski}}]{SPT:Tandoi:2024}
{Tandoi}, C., {Guns}, S., {Foster}, A., {et~al.} 2024, \apj, 972, 6,
  \dodoi{10.3847/1538-4357/ad58db}

\bibitem[{{Thornton} {et~al.}(2016){Thornton}, {Ade}, {Aiola}, {Angil{\`e}},
  {Amiri}, {Beall}, {Becker}, {Cho}, {Choi}, {Corlies}, {Coughlin}, {Datta},
  {Devlin}, {Dicker}, {D{\"u}nner}, {Fowler}, {Fox}, {Gallardo}, {Gao},
  {Grace}, {Halpern}, {Hasselfield}, {Henderson}, {Hilton}, {Hincks}, {Ho},
  {Hubmayr}, {Irwin}, {Klein}, {Koopman}, {Li}, {Louis}, {Lungu}, {Maurin},
  {McMahon}, {Munson}, {Naess}, {Nati}, {Newburgh}, {Nibarger}, {Niemack},
  {Niraula}, {Nolta}, {Page}, {Pappas}, {Schillaci}, {Schmitt}, {Sehgal},
  {Sievers}, {Simon}, {Staggs}, {Tucker}, {Uehara}, {van Lanen}, {Ward}, \&
  {Wollack}}]{ACT:Thornton:2016}
{Thornton}, R.~J., {Ade}, P.~A.~R., {Aiola}, S., {et~al.} 2016, \apjs, 227, 21,
  \dodoi{10.3847/1538-4365/227/2/21}

\bibitem[{{Torne}(2017)}]{Torne:2017b}
{Torne}, P. 2017, PhD thesis, Max-Planck-Institute for Radioastronomy, Bonn

\bibitem[{{Torne} {et~al.}(2015){Torne}, {Eatough}, {Karuppusamy}, {Kramer},
  {Paubert}, {Klein}, {Desvignes}, {Champion}, {Wiesemeyer}, {Kramer},
  {Spitler}, {Thum}, {Gusten}, {Schuster}, \& {Cognard}}]{Torne:2015}
{Torne}, P., {Eatough}, R.~P., {Karuppusamy}, R., {et~al.} 2015, \mnras, 451,
  L50, \dodoi{10.1093/mnrasl/slv063}

\bibitem[{{Torne} {et~al.}(2017){Torne}, {Desvignes}, {Eatough}, {Karuppusamy},
  {Paubert}, {Kramer}, {Cognard}, {Champion}, \& {Spitler}}]{Torne:2017}
{Torne}, P., {Desvignes}, G., {Eatough}, R.~P., {et~al.} 2017, \mnras, 465,
  242, \dodoi{10.1093/mnras/stw2757}

\bibitem[{{Torne} {et~al.}(2020){Torne}, {Mac{\'\i}as-P{\'e}rez}, {Ladjelate},
  {Ritacco}, {S{\'a}nchez-Portal}, {Berta}, {Paubert}, {Calvo}, {Desvignes},
  {Karuppusamy}, {Navarro}, {John}, {S{\'a}nchez}, {Pe{\~n}alver}, {Kramer}, \&
  {Schuster}}]{Torne:2020}
{Torne}, P., {Mac{\'\i}as-P{\'e}rez}, J., {Ladjelate}, B., {et~al.} 2020, \aap,
  640, L2, \dodoi{10.1051/0004-6361/202038504}

\bibitem[{{Torne} {et~al.}(2021){Torne}, {Desvignes}, {Eatough}, {Kramer},
  {Karuppusamy}, {Liu}, {Noutsos}, {Wharton}, {Kramer}, {Navarro}, {Paubert},
  {Sanchez}, {Sanchez-Portal}, {Schuster}, {Falcke}, \&
  {Rezzolla}}]{Torne:2021}
{Torne}, P., {Desvignes}, G., {Eatough}, R.~P., {et~al.} 2021, \aap, 650, A95,
  \dodoi{10.1051/0004-6361/202140775}

\bibitem[{{Torne} {et~al.}(2022){Torne}, {Bell}, {Bintley}, {Desvignes},
  {Berry}, {Dempsey}, {Ho}, {Parsons}, {Eatough}, {Karuppusamy}, {Kramer},
  {Kramer}, {Liu}, {Paubert}, {Sanchez-Portal}, \& {Schuster}}]{Torne:2022}
{Torne}, P., {Bell}, G.~S., {Bintley}, D., {et~al.} 2022, \apjl, 925, L17,
  \dodoi{10.3847/2041-8213/ac4caa}

\bibitem[{{Tziamtzis} {et~al.}(2009){Tziamtzis}, {Lundqvist}, \&
  {Djupvik}}]{Tziamtzis:2009}
{Tziamtzis}, A., {Lundqvist}, P., \& {Djupvik}, A.~A. 2009, \aap, 508, 221,
  \dodoi{10.1051/0004-6361/200912031}

\bibitem[{Weltevrede \& Johnston(2008)}]{Weltevrede:2008:profile}
Weltevrede, P., \& Johnston, S. 2008, Monthly Notices of the Royal Astronomical
  Society, 391, 1210

\bibitem[{{Whitehorn} {et~al.}(2016){Whitehorn}, {Natoli}, {Ade}, {Austermann},
  {Beall}, {Bender}, {Benson}, {Bleem}, {Carlstrom}, {Chang}, {Chiang}, {Cho},
  {Citron}, {Crawford}, {Crites}, {de Haan}, {Dobbs}, {Everett}, {Gallicchio},
  {George}, {Gilbert}, {Halverson}, {Harrington}, {Henning}, {Hilton},
  {Holder}, {Holzapfel}, {Hoover}, {Hou}, {Hrubes}, {Huang}, {Hubmayr},
  {Irwin}, {Keisler}, {Knox}, {Lee}, {Leitch}, {Li}, {McMahon}, {Meyer},
  {Mocanu}, {Nibarger}, {Novosad}, {Padin}, {Pryke}, {Reichardt}, {Ruhl},
  {Saliwanchik}, {Sayre}, {Schaffer}, {Smecher}, {Stark}, {Story}, {Tucker},
  {Vanderlinde}, {Vieira}, {Wang}, \& {Yefremenko}}]{SPT:Whitehorn:2016}
{Whitehorn}, N., {Natoli}, T., {Ade}, P.~A.~R., {et~al.} 2016, \apj, 830, 143,
  \dodoi{10.3847/0004-637X/830/2/143}

\bibitem[{{Wielebinski} {et~al.}(1993){Wielebinski}, {Jessner}, {Kramer}, \&
  {Gil}}]{Wielebinski:1993}
{Wielebinski}, R., {Jessner}, A., {Kramer}, M., \& {Gil}, J.~A. 1993, \aap,
  272, L13

\bibitem[{{Yuan} {et~al.}(2016){Yuan}, {Wang}, {Lei}, {Gao}, \&
  {Zhang}}]{Yuan:2016}
{Yuan}, Q., {Wang}, Q.~D., {Lei}, W.-H., {Gao}, H., \& {Zhang}, B. 2016,
  \mnras, 461, 3375, \dodoi{10.1093/mnras/stw1543}

\end{thebibliography}
\bibliographystyle{aasjournal}

\end{document}